\documentclass{aastex}
\makeatletter
\renewcommand{\fps@figure}{thbp}
\renewcommand{\fps@table}{thbp}

\makeatother
\def\la{\mathrel{\hbox{\rlap{\hbox{\lower4pt\hbox{$\sim$}}}\hbox{$<$}}}}
\def\sun{\hbox{$\odot$}}

\begin{document}
\title{Cepheids and Long Period Variables in NGC\,4395}

\author{F. Thim\altaffilmark{1,3}}
\affil{National Optical Astronomy Observatory \\
950 North Cherry Ave., Tucson, AZ 85726} 
\email{thim@noao.edu}

\author{ J. G. Hoessel\altaffilmark{2,3}}
\affil{Washburn Observatory, University of Wisconsin-Madison, 475 N. Charter Street, Madison, Wisconsin 53706}
\email{hoessel@astro.wisc.edu}

\author{A. Saha\altaffilmark{1,2,3}, J. Claver\altaffilmark{1,3}, A. Dolphin\altaffilmark{1,3}} 
\affil{National Optical Astronomy Observatory \\
950 North Cherry Ave., Tucson, AZ 85726} 
\email{saha@noao.edu, jclaver@noao.edu, dolphin@noao.edu}
 
\and
   
\author{G. A. Tammann\altaffilmark{1,3}}
\affil{Astronomisches Institut der Universit\"at Basel \\ 
Venusstrasse 7, CH-4102 Binningen, Switzerland} 
\email{G-A.Tammann@unibas.ch}

\altaffiltext{1}{NOAO is operated by the Association of Universities for
Research in Astronomy, Inc. (AURA) under cooperative agreement 
with the National Science Foundation.}

\altaffiltext{2}{Visiting Astronomer, Kitt Peak National Observatory, National Optical Astronomy Observatory, 
which is operated by the Association of Universities for Research in Astronomy, 
Inc. (AURA) under cooperative agreement with the National Science Foundation.}

\altaffiltext{3}{The WIYN Observatory is a joint facility of the University of Wisconsin-Madison, 
Indiana University, Yale University, and the National Optical Astronomy Observatory.} 

\begin{abstract}
Repeated imaging observations of NGC\,4395
were made
with the WIYN 3.5 m and the KPNO 2.1 m telescopes.
From the photometry of the resolved brighter stars in this galaxy
eleven Cepheids with periods ranging between
12 and 90 days have been identified. 
The true distance modulus has been derived from the apparent
distance moduli in $g$, $r$ and $i$.
The distance modulus is
$28.02 \pm 0.18$ based on the LMC P-L relation by 
\citet{Sandage:etal:2003}; 
this corresponds to a distance of $4.0 \pm 0.3$ Mpc.
Using the P-L relation 
from \citet{Madore:Freedman:1991}, the distance modulus
is $28.15 \pm 0.18$; which
corresponds  to a distance of $4.3 \pm 0.4$ Mpc.
The reddening is calculated to be $E(g-r)$ = $0.06 
\pm 0.08$ and $E(r-i)$ = $0.10 \pm 0.08$,
again from the distance moduli $\mu_g$, 
$\mu_r$ and $\mu_i$.
In addition, 37 other
variables have been detected, the majority of which have 
definite periods. They are probably all red long period 
variables. 
\end{abstract}


% ******************************************************************
% 1. Introduction 
% ******************************************************************
\section{Introduction}

NGC~4395 ($\alpha_{2000}$=$12^{\rm h}25^{\rm m}48.92^{\rm s}$, 
$\delta = +33^{\circ}32^{\rm m}48.3^{\rm s}$) is a
face-on spiral galaxy that is
classified as Sd III-IV in the Carnegie Atlas of Galaxies
\citep{Sandage:Bedke:1994}. 
It is located in the Ursa Major region called Group B4
by \citet{Kraan-Korteweg:Tammann:1979} and CVn I by 
\citet{de Vaucouleurs:1975}.

NGC~4395 is a pure disk galaxy, i.e. it lacks a bulge
\citep{Filippenko:Ho:2003}. 
It is also the nearest known Seyfert 1 galaxy and has
a central black hole for which the total 
mass has recently been derived by 
\citet{Filippenko:Ho:2003}. They estimate the
mass of the central black hole to be 
$\sim$ 10$^{4}$ $-$ 10$^{5}$ $M_{\sun}$, which is less 
massive than those which have been found in other AGNs.
All other galaxies with known supermassive black holes in their
centers have bulges, which makes NGC~4395
a unique object. 

Literature distance estimates range between 
2.6 Mpc \citep{Rowan:Robinson:1985}
and 5.0 Mpc \citep{Sandage:Bedke:1994}. More recently,
\cite{Karachentsev:Drozdovsky:1998} 
estimated the distance to NGC~4395 
to be 4.2 $\pm$ 0.8 Mpc using the brightest blue stars.
Hubble Space Telescope/WFPC2 images have been used by two groups
for the distance estimate, 
\cite{Karachentsev:etal:2003} obtained a tip of the red giant branch (TRGB) 
distance of 4.61 $\pm$ 0.57 Mpc, \cite{Minitti:etal:2003} derived a 
distance of 4.2 Mpc.

Repeated time spaced images of NGC\,4395 were obtained as part of a program
begun by Saha \& Hoessel in 1990 to determine Cepheid distances
to several galaxies in the Local Group as well as beyond. The project was 
motivated in part by the desire to extend the range of ground based 
Cepheid distances 
by employing the technological advantage 
of CCDs. While the {\it Hubble Space Telescope} would surely vastly extend 
this range, we proceeded on the premise that relatively nearby objects which 
can be reached from the ground would not, and should not, occupy the 
precious and expensive resources of the {\it HST}.  Nevertheless, the 
cartography of the local Universe continued to be an important problem.

Ground based time-sequence observations are inevitably impeded by constraints 
from telescope scheduling, weather and poor seeing. As a result, some 
of the objects (particularly the more distant and more demanding ones) 
selected in our initial program with the Kitt Peak 2.1-m telescope 
could not be completed. With the commissioning of the WIYN 3.5-m 
telescope (also on Kitt Peak) we got not only an increase in aperture, but 
also better image quality. Additionally, and very important, 
for a time while the NOAO queue observations were being conducted regularly, 
the necessary observations 
for the unfinished galaxies could be obtained efficiently. 

While the time span to obtain the observations for the Cepheid discovery 
in NGC\,4395 took a frustrating 8 years, a side benefit 
is that many long period variables (LPVs) were also discovered, for 
which periods could be determined. The literature lacks period, luminosity 
and color information for LPVs, except in a very few galaxies (the Magellanic 
Clouds, M33), and consequently the empirical properties of LPVs and their 
dependence on metallicity (for instance) are very poorly known. We hope 
that the LPVs we have been able to catalog here for NGC\,4395 is a step 
towards ameliorating this situation.

% ******************************************************************
% 2. Observations
% ******************************************************************
\section{Observations} 

Images of NGC\,4395 were taken using two different telescopes,
the WIYN 3.5 m telescope, and the 2.1 m KPNO telescope, which
are both located on Kitt Peak. These repeated images were obtained
over a period of 8 years, from 1991 January 13, to
1999 January 13 in the $g$, $r$ and $i$ passbands of the 
\citet{Thuan:Gunn:1976} system. 
 
The detectors which have
been used are named TI-2, TIKA and S2KB. TI-2, used on the 2.1 m,
is a 796 $\times$ 796 CCD
with a pixel size of 15 $\mu$m and a scale of $0\farcs2/$pixel.
T1KA, also used on the 2.1 m, is a  1024 $\times$ 1024
CCD which provides a pixel scale of $0\farcs305/$pixel and a field of view
of $5\farcm2 \times $5\farcm2. S2KB was
exclusively for the WIYN telescope, it is a 2048 $\times$ 2048
CCD with 21 $\mu$m pixels. This provides a pixel scale of 
$0\farcs197/$pixel and a field of view
of $6\farcm8 \times $6\farcm8. 
The exposure times are between 500 and 1200 s.
A journal of the observations is given in Table~\ref{tab1}.

Primary observations were in the $r$-band, 
which was optimal for the color response of the early CCDs used, given both, the 
color range of Cepheids, and the region of the spectrum where the sky is 
sufficiently dark. Supplemental observations were made at four epochs in the 
$g$ band and at two epochs in the $i$ band, 
so that mean colors for the Cepheids could be derived, 
and used to track extinction and reddening. 

% ******************************************************************
% 3. Data Processing
% ******************************************************************

\section{Data Processing} 

\subsection{Relative Photometry and Object Matching}

Multiple exposures obtained with WIYN S2KB were combined 
and used as a template fits file. The reference image in
$r$ is shown in Fig.~\ref{Thim.fig1.ps}. 
This image is also provided as a FITS file in the electronic version of this 
paper. Positions ($X$ and $Y$ in pixels) 
of objects given in various tables in this paper can be easily located 
on the electronic image. The published FITS image is also appointed 
with a `world coordinate system' (WCS) calibrated to J2000 coordinates
on the sky. 

The reference image is used as a
position reference only. For visualization and `blinking' of images, 
the individual epoch images were mapped to the reference image.
However, photometry (see below) for the individual epochs 
was always done on the untransformed original images, so that noise 
characteristics on the image remain untainted, and also to avoid 
the seeing degradation from the transformation.

Relative photometry and object identification on the template
images were obtained with the
program DoPHOT \citep{Schechter:Mateo:Saha:1993}. 
Subsequently DoPHOT was run on each individual epoch. 
All objects
identified in the individual epochs were matched to the
template frame and the coordinate transformations 
for the individual epochs were derived. The relative 
photometry at each epoch differs from that in the template frame by a
constant magnitude offset. This offset is easily determined from the 
ensemble average of the difference between matched stars in the template
and the epoch. In order to avoid issues with bias at faint magnitudes, as well
as with bad outliers, this procedure is best done by inspecting a plot 
of the magnitude difference for each star versus the magnitude on the 
template image.
This puts the relative magnitudes of all stars in all epochs 
in $r$ on the same footing. A similar procedure was used for 
the $g$ and $i$ passbands as well.
In practice, the template magnitudes in the 3 bands had already been 
adjusted to the true Thuan-Gunn system using the procedure described 
in the following sub-section, so the final magnitudes for each epoch
were also on the Thuan-Gunn system. 

\subsection{Calibrating the Photometry}

Additional observations of NGC\,4395 along with observations of
standard stars were taken with 
the S2KB detector on the WIYN 3.5m  telescope 
in the Thuan-Gunn $g$, $r$, $i$ bands on 2000 May 26 (UT date).
The images of NGC\,4395 taken on this night were shallower exposures 
to ensure that some of the brighter isolated stars would not be saturated. 
A photometric solution was obtained for that night from the standard 
star observations. The error in the mean for the primary 
standard stars on this night is less than 0.01 mag in all three passbands.

\begin{figure}[!ht]
\epsscale{1.05}
\plotone{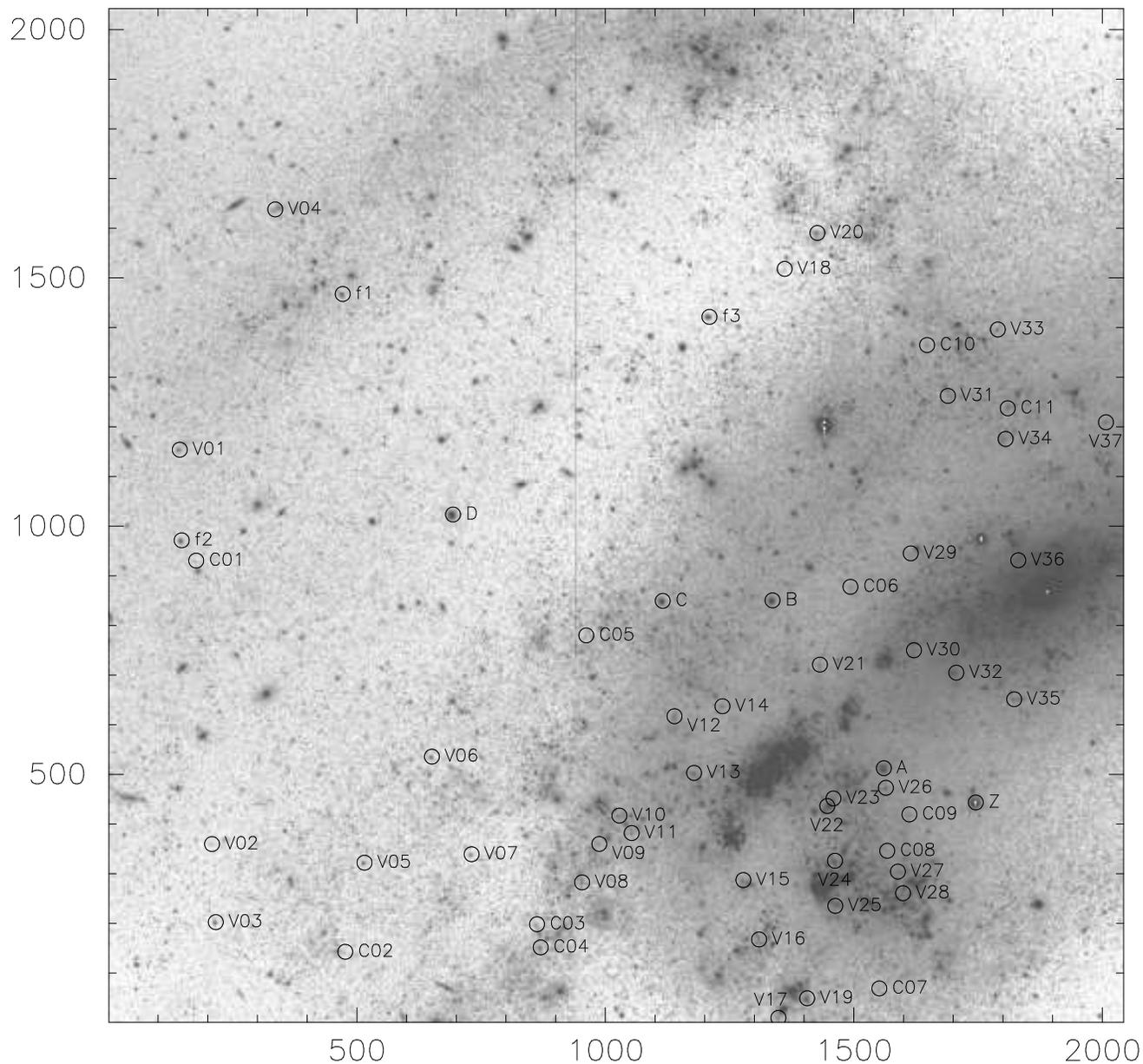}
\vspace{0cm}
\caption{WIYN S2KB $r$ image covering the field NE of the
center of NGC\,4395, which is located on the right side
close to the edge of the chip. North is up and East is to the left.
Cepheids and other long period variables are
shown as open circles and labeled with C for Cepheids and
V for other variables followed by a number. The labels
are the same as given in Table~\ref{tab_cepheids} and 
\ref{tab_variables}. The local standards which are presented
in Table~\ref{tab_stds} are also shown as open circles and are
labeled by the letters A, B, C, D, Z, $f1$, $f2$, and $f3$.}
\label{Thim.fig1.ps}
\end{figure}

Aperture photometry of several bright and relatively 
uncrowded stars in the field of NGC\,4395 were obtained, and true magnitudes 
derived using the photometric solution for that night. Sky subtraction
in all cases was done using the `growth curve' method. 
Thus local standards in $g$, $r$, and $i$ were established in the field 
of NGC\,4395. These local standards are identified by the letters 
A, B, C, D and Z, and are shown in Fig.~\ref{Thim.fig1.ps}. 
Their magnitudes are listed in Table~\ref{tab_stds}
along with their $X$ and $Y$ positions on the template image, and 
also their coordinates on the sky.
The standard error of measurement for each star is less than $0.015$ mag rms. 
In addition, 3 relatively fainter stars were also calibrated as local 
standards, but using PSF fitting and aperture correction. These stars are 
designated as $f1$, $f2$, and $f3$, and are also shown in Fig.~\ref{Thim.fig1.ps}
with magnitudes listed in Table~\ref{tab_stds}. For these 3 fainter objects, 
the standard error is $\sim 0.03$ mag rms per star. 
A conservative estimate is that the magnitude scales and colors are 
calibrated to within a systematic error not exceeding 0.05 mag.

Armed with this calibration, the relative photometry described in the 
preceding section is adjusted to the true Thuan-Gunn system by comparing
the relative photometry of an ensemble of several thousand stars from any epoch
or template with stars found in common on the images taken on the 
calibrating night.

%
%
%
% ******************************************************************
% 4. Identification of the Variable Stars
% ******************************************************************

\section{Identification of the Variable Stars}

\subsection{Detection of Variable Stars}

Given photometry and photometric errors in all epochs, each 
object was examined for variability. The method for
identifying variable stars is described in \citet{Saha:Hoessel:1990}
and is not repeated here.
The requirements for variability have been improved since
\citet{Saha:Hoessel:1990} from experience with similar data. 
The individual magnitudes measured in the various epochs 
and their individual errors estimates for
each of the Cepheids in $g$, $r$ and $i$ are tabulated in 
Table~\ref{tab_phot_cep}, the individual magnitudes and 
errors of the other variable stars
in $g$, $r$ and $i$ are tabulated in Table~\ref{tab_phot_var}.

If $P$ is the
period of a supposed variable star, $m_i$ the measured magnitude at the
ith epoch, and $\overline{m}$ the average over the n values of $m_i$, and 
if the values for $m_i$ are arranged in increasing order of phase, 
then $\Theta$ is defined as: 
\begin{eqnarray}
\label{def_theta}
\Theta(P) \; = \; \frac{\sum_{i=1}^n(m_{i + 1} \; - m_i)^2}{\sum_{i=1}^n
(m_i \; - \overline{m})^2}.
\end{eqnarray}
A minimum in the spectrum of $\Theta$ indicates a possible period. 
The reduced $\chi^2$ is given by:
\begin{eqnarray}
\label{def_chi_sq}
\chi^2_{\nu} \; = \; \sum_{i}^n\frac{(m \; - \overline{m})^2}{\nu \, {\sigma^2_i}},
\end{eqnarray}
where $\sigma_i$ is the rms error
and $\nu$ = n - 1.
$\Lambda$ \citep{Lafler:Kinman:1965, Saha:Hoessel:1990}
is related to $\Theta$ in the sense that 
$\Lambda$ goes as $\frac{1}{\Theta}$. 
An object is considered to be variable if the
$\chi^2$ probability is higher than 0.99 and $\Lambda$ greater than 
3.5. Furthermore, an object with a reduced $\chi^2$ higher than 100
is always flagged to be a possible variable.
The requirements and changes from \citet{Saha:Hoessel:1990}
are described in more detail 
in \citet{Saha:Claver:Hoessel:2002}.  

Sixty four out of a total
of 25,155 stars in the template $r$ frame are flagged to be 
possible variables.
These candidates are examined for periodicity using
the method of \cite{Lafler:Kinman:1965}. The period range from
1 to 3000 days has been used to probe possible variability.
Only data in the $r$ band have been used for the variable search 
because only 4 epochs in the $g$ and 2 in the $i$ band have been taken,
which is insufficient to derive a period. 
The objects which are
likely to be Cepheids are presented in Table~\ref{tab_cepheids}, 
the remaining other variables are listed in 
Table~\ref{tab_variables}. As a double check on 
the variability all
candidates have been visually blinked
and compared with non-variable stars. 
The periods are given in column 2.
The periods for some variables in Table~\ref{tab_variables}
are uncertain, with uncertainties ranging from 
a few days to a few hundred days. In such
cases, we present our best guess in the table.

The mean magnitudes in $r$ are phase-weighted intensity 
averaged means and were calculated using the
following equation:
\begin{eqnarray}
\label{phase_weighted_mean}
<\!r\!> \; = \; - 2.5 \; \log_{10} \sum_{i=1}^{n} \; 0.5 \; ( \phi_{i+1} - \,\phi_{i-1} ) \, 
10^{-0.4 \,r_i}, 
\end{eqnarray}
where 
n is the number of observations, $r_i$ the magnitude, and $\phi_i$ 
the phase of the ith observation in order of increasing phase. Intensity
weighted magnitudes can be biased due to missing measurements,
-- biased in a sense that it is more likely to detect the variable
star at a brighter than at a fainter phase.
The phase-weighted intensity mean gives isolated points more
weight than closely spaced ones, which makes it superior to a straight
intensity mean. 
The method to use well-sampled lightcurves to predict corrections
to small samplings at another wavelength in order to predict their 
mean magnitudes originates with \citet{Freedman:1988}. We are using 
here a similar method described in \cite*{Labhardt:etal:1997}.
The mean $g$ magnitudes in 
Table~\ref{tab_cepheids} are calculated using the information
on the shape and the amplitude of the complete  
light curve in one filter as well as the typical phase shift between 
two filters are used to derive a value of $<\!g\!>$ from each
$g$ measurement.
\cite{Labhardt:etal:1997} presents transformations
for the filters B, V, R and I. Therefore, the empirical correction
function between V and R have been transformed to $g$ and $r$
using synthetic transformations and amplitude ratios. 

\subsection{Identification of Cepheid Variables and Period Determination}

\begin{figure}[!ht]
\epsscale{0.75}
\plotone{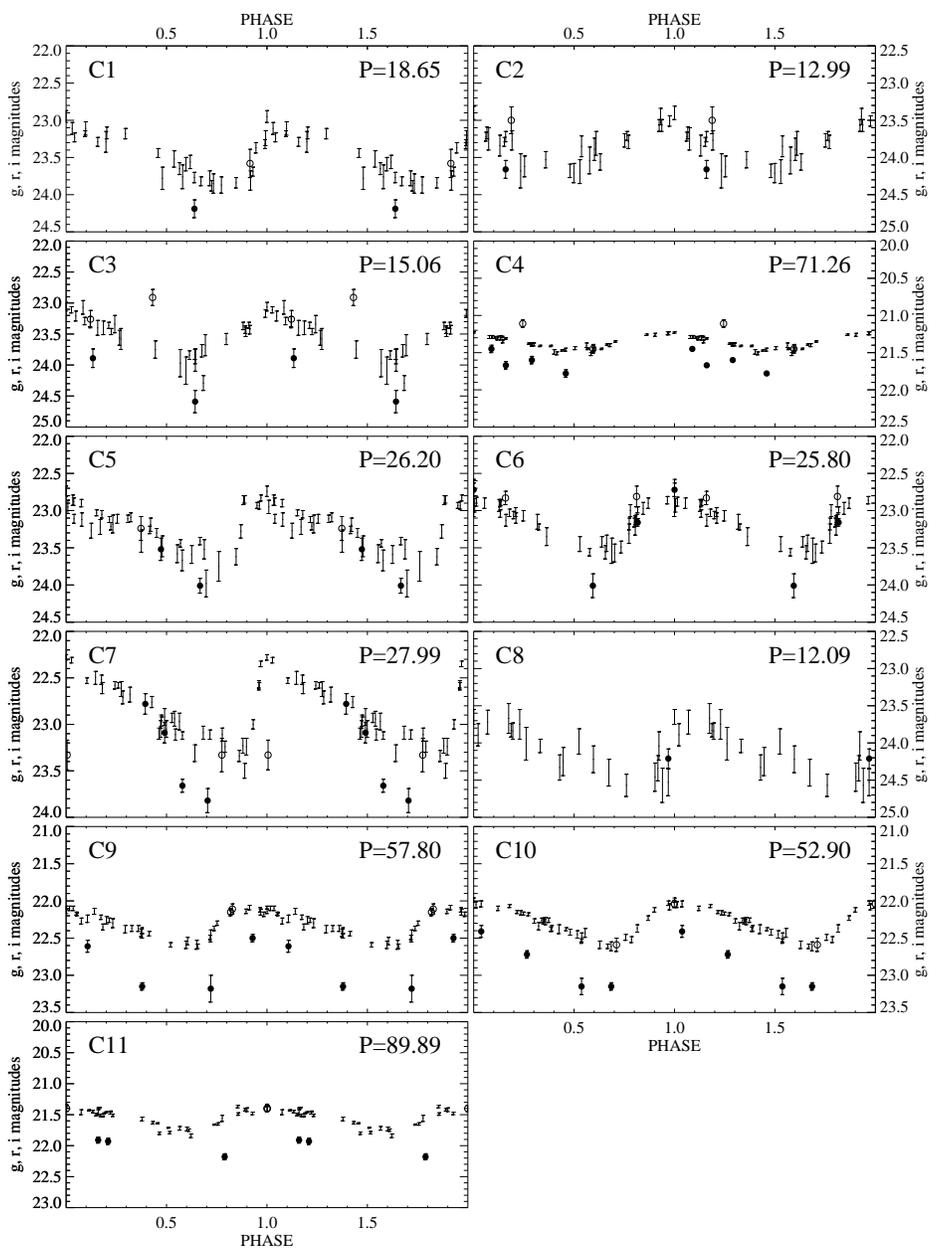}
\vspace{0cm}
\caption{The light curves of the Cepheids in $r$ (small 
dots with error bars), 
in $g$ (large filled circles) and in $i$ (large open circles). 
The identification numbers (C1-C11) and its periods
are displayed in each graph. The periods
are given in days.}
\label{Thim.fig2.eps}
\end{figure}

The light curves of all possible variable stars with 
reasonable values of $\Theta$ were individually inspected by
eye in all three passbands. However, in $g$ and $i$ there
have been sometimes only one or even zero photometric
measurements, which makes the term ``light curve''
meaningless. In such cases, the color estimate is rather
uncertain.
The light curves of the Cepheid candidates are shown in 
Fig.~\ref{Thim.fig2.eps}.

The quality of the light curve in $r$, the quality of the light curve in $g$,
the phase 
coherence between the $r$, the $g$ and $i$ light curves, the shape
of the spectrum of $\Theta$ and the $g-r$ and $r-i$
colors have been used as the selection 
criteria to consider a variable star being a Cepheid or not.

Since there are very few epochs with measurements in the $g$ band, 
the mean $<\!g\!>$ magnitudes are calculated using the 
method of \cite{Labhardt:etal:1997}, with an equation of the form: 
\begin{eqnarray}
\label{eq_labhardt_mean}
<\!g\!>\;\; =&&\!\ g(\phi_r)  +  [<\!r\!> - r(\phi_r)] + \Delta r \,\, C_{r \rightarrow g}(\phi), 
\end{eqnarray}
where $\Delta r$ is the $r$ amplitude, $<\!r\!>$ the phase-weighted 
mean $r$ magnitude,
$\phi$ the phase of the light curve and $C_{r\rightarrow g}(\phi)$ the
empirical function for the transformation between $r$ and $g$ 
magnitudes. 
\cite{Labhardt:etal:1997} listed values of $C(\phi)$ for $B,V,R,I$ mags, 
from which we can derive $C_{r\rightarrow g}(\phi)$, using the equation:
\begin{eqnarray}
\label{c_r_g}
C_{r\rightarrow g}(\phi) = 1.722 C_{V\rightarrow R}(\phi) + 0.655 C_{V\rightarrow B}(\phi),
\end{eqnarray}
$C_{V\rightarrow R}(\phi)$ are the correction values for $V$ $\rightarrow$
$R$ magnitudes, $C_{V\rightarrow B}(\phi)$ the value
for $V$ $\rightarrow$ $B$ magnitudes, and are listed in 
\cite{Labhardt:etal:1997}. The derivation of the above equation uses 
color transformations that are discussed in \S~\ref{par5}. 

The mean of the individual $<\!g\!>$ magnitudes yields the adopted
value of $<\!g\!>$ and its error. 
The $i$ magnitudes are calculated from the $r-i$ color at the
$i$ phase and the mean $<\!r\!>$ values, (i.e. setting 
$C_{r\rightarrow i}(\phi)$ to zero).

In total, we identified 11 Cepheids, which are listed in 
Table~\ref{tab_cepheids}. Column~1 gives the designation
of the Cepheid, column~2 the period, columns~3\,-\,8 
give the mean magnitudes and errors in $g$, $r$, and $i$, and columns~9\,-\,10
their position on the template $r$ image. 

As noted before, 
all variable stars have been visually blinked. Nine of the eleven Cepheids
in Table~\ref{tab_cepheids} blink, whereas it is hard to tell
for the Cepheids C4 and C11 because of their small amplitudes.
Besides these eleven candidates no further Cepheid candidates have been 
accepted. 

\subsection{Long Period Variables}

The remaining candidates, which have not been classified as
Cepheid variables, are listed in Table~\ref{tab_variables}.
We found 37 variables besides the 11 Cepheids.
Again, column~1 gives the designation
of the variable, column~2 the period, columns~3\,-\,5 
give the mean magnitudes in $g$, $r$, and $i$, 
column~6 the $r$ peak to peak amplitude, 
and columns~7\,-\,8
their position on the template $r$ image. 
The $r$ amplitude is estimated
from the peak to peak magnitude from their smoothed light curves rather than 
from the difference of minimum and maximum measured magnitudes.
All variables have been
visually inspected by blinking pairs of images in order to see
variability, i.e. images near maximum brightness were
compared against images near minimum brightness. All 
candidates in Table~\ref{tab_variables} are varying intrinsically
in brightness. 16 of these variables have an uncertain period; 
in these cases the best guess for the period estimate is
presented. Their light curves are presented in 
Fig.~\ref{Thim.fig3a.eps}. 

The variables in Fig.~\ref{Thim.fig6.eps} have $g-r$ colors
between 0.5 and 1.6 and $r-i$ colors of about 0.2 to 1.2,
which indicates that these variables are red giants. 
No Luminous Blue Variable (LBVs), or Hubble-Sandage Variable 
\citep{Hubble:Sandage:1953} has been identified.
A detailed discussion of the LPVs is taken up in
\S \ref{disc_LPVs} later in the paper after we have derived 
the distance modulus.

% ******************************************************************
% 5. The Period-Luminosity Relation 
% ******************************************************************

\section{The Cepheid Period-Luminosity Relation} 
\label{par5}
\subsection{The Period-Luminosity Relation using the Kent Transformations}

The transformation 
equations from the Johnson to the Thuan-Gunn filter system
given by \cite{Kent:1985} are:
\begin{equation}
\label{eq_g}
V \; =  \; g - \; 0.03 \; -0.42 \; (g-r) 
\end{equation}
\begin{equation}
\label{eq_r}
R \; =  \; r - \; 0.51 \; -0.15 \; (g-r).
\end{equation}

For the $i$ band the \citet{Wade:etal:1979} transformations have been used.
These transformations which are valid for a color range which are typical
for Cepheids are given by:
\begin{equation}
\label{eq_i}
i \; =  \; 0.999 I + \; 0.690 \; +0.419 \; (R-I). 
\end{equation}

The P-L relation in $BVRI$  
from \citet{Madore:Freedman:1991}, can be
transformed using the relations by \cite{Kent:1985} and
\citet{Wade:etal:1979} to be:

\begin{equation}
\label{eq:PL_M_F_g}
 M_{g} ~~=~~ -2.62 \, (\log P - 1) - 4.08~, 
\end{equation} 
\begin{equation}
\label{eq:PL_M_F_r}
 M_{r} ~~=~~ -2.91 \, (\log P - 1) - 4.04~,  
\end{equation} 
\begin{equation}
\label{eq:PL_M_F_i}
 M_{i} ~~=~~ -3.00 \, (\log P - 1) - 4.06~,  
\end{equation} 

as derived in \citet{Hoessel:etal:1994}.

\subsection{The Period-Luminosity Relation using Synthetic Transformations}
\label{par5.2}

Transformations that better reflect the spectral energy distribution
(SED's) of supergiants with temperatures spanning that of Cepheids
can be obtained synthetically.

The observed SED's of several Thuan-Gunn standard stars
spanning a wide range of color integrated for model
bandpasses of $g$, $r$ and $i$. The transformation to convert
the resulting 'instrumental' magnitudes to true magnitudes
on the Thuan-Gunn system was derived. 

The same was done for the $BVRI$ Johnson-Cousins-Landolt system.

Once we were able to accurately reproduce the stellar photometry from the
stellar spectra, we computed synthetic magnitudes in both filter systems
-- Johnson-Cousins-Landolt and Thuan-Gunn -- 
from synthetic stellar spectra \citep{Kurucz}.
Limiting this work to stars with temperatures and gravities within and
slightly beyond the range of Cepheids, we computed the
following transformations from the Johnson-Cousins-Landolt 
system to the Thuan-Gunn system:
\begin{eqnarray}
\label{andy1}
  g-V \, = \, -0.102 \, + \, 0.393 \, (B-V)
\end{eqnarray}
\begin{eqnarray}
\label{andy2}
  r-R \, = \ 0.437 \, - \, 0.033 \, (V-R)
\end{eqnarray}
\begin{eqnarray}
\label{andy3}
  i-I \, = \, 0.816 \, - \, 0.081 \, (R-I).
\end{eqnarray}
Because these transformations were computed over a limited range of
temperature, gravity, and metallicity, no second-order color terms are
necessary. 
%to reduce the scatter in the fits. 
Naturally, this also means
that these transformations are useful only for stars in this restricted
range of parameters.
Because the range in color of the Cepheids under consideration
is quite limited, the linear approximation in 
eq.~\ref{andy1}-\ref{andy3} is sufficient.

Applying these synthetic transformations to the P-L relations of \citet{Madore:Freedman:1991} 
leads to the P-L relations in the Gunn system:
\begin{eqnarray}
\label{andy4}
  M_g \, = \, -2.63 \, (\log P-1) \, - \, 4.00~,
\end{eqnarray}
\begin{eqnarray}
\label{andy5}
  M_r \, = \, -2.95 \, (\log P-1) \, - \, 4.09~,
\end{eqnarray}
\begin{eqnarray}
\label{andy6}
  M_i \, = \, -3.07 \, (\log P-1) \, - \, 4.08.
\end{eqnarray}

Eq.~\ref{andy5} $-$ \ref{andy6} ($r$ and $i$ bands) compare 
favorably with their counterparts (equations~\ref{eq:PL_M_F_r} and 
\ref{eq:PL_M_F_i})
from \citet{Hoessel:etal:1994} based on the \citet{Kent:1985} transformations. 
But in the $g$ band (eq.~\ref{andy4} compared to eq.~\ref{eq:PL_M_F_g})
the difference is quite significant. 
The effect in terms
of measured apparent distance moduli is:
\begin{eqnarray}
\label{andy7}
  \delta \, \mu_g \, = \, 0.01\  (\log P-1) \ - \ 0.08
\end{eqnarray}
\begin{eqnarray}
\label{andy8}
  \delta \ \mu_r \, = \, 0.04 \, (\log P-1) \ + \ 0.05
\end{eqnarray}
\begin{eqnarray}
\label{andy9}
  \delta \, \mu_i \ = \, 0.07 \ (\log P-1) \ + \ 0.02,
\end{eqnarray}
where the sign is distance modulus from eq.~\ref{andy4} $-$ \ref{andy6}
minus \citet{Hoessel:etal:1994}. 
For a Cepheid with a period of
25 days, the differences are $-0.08$ magnitudes in $g$, +0.07 magnitudes in $r$,
and +0.04 magnitudes in $i$. 
The synthetic transformations are superior to the older ones,
since they are based on empirical observations of stars 
which are mostly dwarfs, and whose SEDs differ systematically 
from supergiants.
Our previous studies were based on $ri$
photometry, and thus the color errors were small enough to be  
unnoticeable -- but here, when the $g$ band is used, the
synthetic transformations are necessary, particularly
since we will see in \S~\ref{chap_6} that the Kent transformations
imply reddening values that are inconsistent with the
observed colors of the bluest stars in NGC~4395.

The observed P-L relations are shown in Fig.~\ref{Thim.fig4.eps} for
the $g$ and the $r$ bands. Since the slope is fixed, the 
remaining free parameter is the offset which corresponds to 
the apparent distance moduli in $g$ and $r$. The dashed lines
indicate the intrinsic width of the P-L relation. The reported
uncertainties are the standard errors, i.e. the rms scatter
divided by the square root of the number of Cepheids.

% ******************************************************************
% 6. Extinction Estimate from Cepheids and The Distance Modulus 
% ******************************************************************
\section{Extinction and the Distance Modulus} 
\label{chap_6}

In order to derive the absorption-reddening relations
\begin{eqnarray}
\label{eq:reddening_V}
R_{g,r} =  \frac{A_{g,r}}{E(g-r)}, 
\end{eqnarray}
we have synthesized '$g$' and '$r$' magnitudes for
Cepheid spectra as in section~\ref{par5.2}, but now
additionally including extinction using the 
equations of \citet{Cardelli:Clayton:Mathis:1989}. 
We adjusted the value of $R_V$ to be that for which 
$A_{V}/E(B-V)$ is 3.1.  

By comparing the results from different amounts of input
absorption, we derived $A_{g}/E(g-r)$ = 3.44.
The true distance modulus is then given by

\begin{equation}
\label{mu_0}
\mu_0 \, = \, \mu_g - A(g) \, = \, 3.44 \, \mu_r \, - \, 2.44 \, \mu_g.
\end{equation}
\citep[cf. ][eq. 40.]{Tammann:Sandage:Reindl:2003} 

We have applied the transformed P-L$_{g,r}$ relations
in eq. 15 and 16, which are based on the old P-L$_{B,V,R}$
relations by \citet{Madore:Freedman:1991},
to the 11 Cepheids in NGC~4395. The resulting
apparent distance moduli $\mu_g$ and $\mu_r$ are shown for
each star in Table~\ref{tab_individual} as well as the individual
true distance moduli from eq.~\ref{mu_0}.  
The mean true distance modulus
of NGC~4395 is then found to be
\begin{equation}
\mu_0 \, = \, 28.22 \pm 0.21.
\end{equation} 
The apparent moduli $\mu_g$ and $\mu_r$ in Table~\ref{tab_individual} imply
a mean reddening of the Cepheids of E(g$-$r) = 0.059.

If we use the P-L relation of \cite{Madore:Freedman:1991} with 
transformations by \cite{Kent:1985} the true distance modulus $\mu_0$
would be 27.80 with a large reddening E(g-r) of 0.20  
(see Table~\ref{tab_individual}). 
An examination of our CMD (Fig.~\ref{Thim.fig6.eps})
indicates that the bluest stars are too blue to
have a reddening of $E(g-r)$=0.20, as the blue edge of the CMD is roughly
0.1 magnitudes redward of the theoretical limit of $g-r$=$-0.8$,
which therefore means that $E(g-r)$ cannot be $>$ 0.1.
This convinces us that the P-L relations from the synthetic 
transformations are more
accurate and we adopt them in this paper.

The true distance modulus to NGC~4395 and the reddening can also be
obtained with the mean $r$ and $i$ magnitudes of the Cepheids. The
individual mean magnitudes are listed in Table~\ref{tab_cepheids}.
As shown in Table~\ref{tab_phot_cep}, the number of observations in the
$i$ band is 1 or 2, except C08, which has no $i$ magnitude.
The mean $i$ magnitude is calculated from the $r-i$ color at 
$i$ phase. Using again the P-L relation of \cite{Madore:Freedman:1991} 
and the transformations by \cite{Kent:1985} for the $r$ band and in
addition the transformations by \citet{Wade:etal:1979} for the $i$ band,
we obtain $\mu_i \, = \, 28.20 \pm 0.12$ and $E(r-i) = 0.09 \pm 0.08$.
If we use the P-L relation of \cite{Madore:Freedman:1991} 
with the synthetic transformations the mean 
$\mu_i$ is calculated to be $28.26 \pm 0.13$ 
with a mean $E(r-i)$ reddening of $0.10 \pm 0.09$. 
The individual values are presented in Table~\ref{tab_individual_r_i}.
The true distance modulus is given by
\begin{equation}
\label{mu_0_r_i}
\mu_0 \, = \, 3.72 \, \mu_i \, - \, 2.72 \, \mu_r
\end{equation}
\citep*[][eq. 10]{Saha:Claver:Hoessel:2002}, which leads
to  $\mu_0 \, = \, 27.96 \pm 0.34$ and $27.98 \pm 0.34$ 
for the Kent and Wade and the synthetic transformations,
respectively. 
This agreement demonstrates that use of the Kent-Wade 
transformations for $r$ and $i$ used in previous papers 
is consistent with the new synthetic relations. The estimated
error in the present case, from using $r$ and $i$ is much
larger than that from using $g$ and $r$ because of the fewer 
$i$ observations compared to $g$ observations.
In the following, we exclusively
use the distances obtained with the synthetic transformations.
We calculate the assumed true distance to NGC~4395 with the weighted
mean of the true distances obtained with $g$ and $r$ on one hand, 
and the distance obtained with $r$ and $i$ on the other hand.
The true weighted distance modulus is then given by
combining both sets of distances:
\begin{equation}
\label{mu_0_r_i_number}
\mu_0 \, = \, 28.15 \pm 0.18
\end{equation}
using the \cite{Madore:Freedman:1991} P-L relations.

The Galactic reddening of $E(g-r) < 0.02$ 
\citep{Burstein:Heiles:1984,Schlegel:Finkbeiner:Davis:1998} cannot 
account for the full reddening of the Cepheids. They must
suffer additional reddening in the parent galaxy. The diagnostic
diagram in Fig.~\ref{Thim.fig5.eps} suggest that the intrinsic 
reddening is quite uniform.

A new fundamental problem with Cepheid distances arises from
the fact that P-L relations in different galaxies are
different \citep{Tammann:etal:2002}. The Galactic P-L
relation is markedly steeper than in LMC, the latter beeing in addition
non-linear \citep*{Tammann:Sandage:Reindl:2003}.
The authors show
that at least part of these differences are due to metallicity 
differences. If metallicity alone decides about the form 
of the P-L relation, NGC~4395 with 12 + log O/H = 8.33 +/- 0.25
\citep{Roy:etal:1996} is similar to LMC and should
be compared with the LMC P-L relation. The latter has been derived 
by \citet*{Sandage:etal:2003} from the 
vast data by \citet{Udalski:etal:1999} which are 
augmented from external sources for longer-period Cepheids.
The new LMC P-L relations for Cepheids with periods more than 10 days
in $V$ and $I$ are given by
\begin{equation}
 M_{V} ~~=~~ -(2.609 \pm 0.099) ~\log P - (1.565 \pm 0.131)~, 
\end{equation} 
\begin{equation}
 M_{I} ~~=~~ -(2.864 \pm 0.082) ~\log P - (2.010 \pm 0.108)~.  
\end{equation} 
Therefore, the true final dereddened distance modulus becomes
\begin{equation}
\mu_0 \, = \, 28.02 \pm 0.18.
\end{equation} 
We prefer this value over $\mu_0$ = 28.26 $\pm$ 0.18
which one would obtain from the P-L relation of the Galaxy,
which is clearly more metal-rich than 
NGC~4395.

% ******************************************************************
% 7. The Color-Magnitude Diagram and the LPVs
% ******************************************************************
\section{The Color-Magnitude Diagram, Extinction Estimates and LPVs}

\subsection{The Color-Magnitude Diagram and Extinction Estimates}
\label{chap7.1}

The color-magnitude diagrams for the template frames for $g$, $r$
and $r$, $i$ are shown in Fig.~\ref{Thim.fig6.eps} and 
\ref{Thim.fig7.eps}. The large filled circles mark the position
of the mean magnitudes of the Cepheids, open large circles 
indicate the position of the mean magnitudes of all other
variable stars. The position of the Cepheids
in the CMDs is consistent with expectation.
The majority of other variables have red colors and periods
longer than hundreds of days. 
They are listed in Table~\ref{tab_variables}, their light curves
are presented in Fig.~\ref{Thim.fig3a.eps}.

As stated before,
an examination of our CMD indicates that the bluest stars are too blue to
have a reddening of $E(g-r)$=0.20, as the blue edge of the CMD is roughly
0.1 magnitudes redward of the theoretical limit of $g-r$=$-0.8$. Thus we
believe the P-L relations from the synthetic transformations to be more
accurate and adopt them in this paper.

\subsection{The Long Period Variables}
\label{disc_LPVs}

\citet{Kholopov:etal:1985} defined three types of 
variable red giant stars. First, the Mira type variables are long
period variables with periods in the range 
between 80 and 1000 days and amplitudes in the $V$ band
between 2.5 and 11 mag.
Second, the semiregular variables which have smaller amplitudes, from several 
hundreths to 1-2 mag in the $V$ band, their period range
is between 20 to 2000 days or more and their light 
curves show a less regular behavior. The 
third group of variables are the irregular variables
having amplitudes of the order of 1 mag in the $V$ band.
\citet{Kholopov:etal:1985} also mention that 
many stars are mis-classified as irregular variables because
of incomplete studies.

The \citet{Kholopov:etal:1985} classification is based in the main on 
stars observed within our Galaxy. The selection effects for Galactic 
variable stars is very different from that for external galaxies. 
Within our Galaxy, fainter relatively nearby variables abound, whereas 
very luminous variables, which are also short lived, are relatively rare, 
and since they occur in the disk, are often obscured behind a lot of dust.
On the other hand, in external galaxies, these are much more easily 
seen, since they are among the brightest stars. In a near face-on case 
like in NGC\,4395, extinction and obscuration are less significant.
A better template for comparing the LPVs is from the sample from the 
\citet{Kinman:Mould:Wood:1987} study in M\,33.
In their analysis, \citet*{Kinman:Mould:Wood:1987} identified red long period 
variables (LPVs) as either core-helium-burning as or as upper
asymptotic giant branch (AGB) stars. Core-helium-burning, or red 
supergiant (RSG) stars are
at least 1 magnitude brighter than AGB stars \citep{Wood:Bessell:Fox:1983}.
The RSG long period variables are seen to have smaller amplitudes than
the AGB stars \citep{Wood:Bessel:1985}. In the $r$-band, typical amplitudes 
for RSG variables are a magnitude or smaller, whereas for the AGB LPVs
amplitudes as large as 3 mags are quite common.
Further, an AGB star has a degenerate core, and its luminosity is 
determined by its 
core mass, which in turn is subject to the Chandrasekhar limit. This implies 
an upper limit on the luminosity of an AGB star of $M_{bol} \approx -7.0$, 
which is equivalent to $M_{r} = -5.20$ mag \citep*{Kinman:Mould:Wood:1987}.

To see if the above characteristic criteria can be used 
to distinguish between AGB and RSG LPVs, consider  
Fig.~\ref{Thim.fig8.eps}, where the mean $r$ magnitudes of all possible 
long period variables are plotted against their $r$ amplitude.
A clear break in the $<r>$ magnitude distribution can be seen: there are four 
LPVs fainter than $<r> = 23.2$ with the rest mostly brighter than 
$<r> = 22$ mag. 
Anticipating (in \S \ref{chap_6}) the apparent distance modulus in $r$ 
of $ 28.36 $, we conclude that an AGB star can be no brighter 
than $r = 23.16$. We therefore argue that the four LPVs in 
Fig.~\ref{Thim.fig8.eps} that are fainter than $r = 23.1$ are AGB stars, 
whereas the remainder are RSG LPVs. This argument is borne out by 
the location of these stars on the 
color-magnitude diagram obtained from our image data, 
as discussed in \S \ref{chap7.1}. 

However, the amplitudes of these two `classes' of LPVs do not follow 
expected behavior. The variables deduced to be AGB stars 
all have $r$ amplitudes near 1.0 mag,
which is smaller than typical. All but four of the putative RSG variables
have reasonably small amplitudes, but the four with $\Delta r > 1.2$ mag 
clearly stand out in Fig.~\ref{Thim.fig8.eps}. This is not entirely without 
precedent, since \citet{HSD:1998} found three very high amplitude red 
variables in the dwarf galaxy DDO\,187, where these stars are among the 
brightest stars in that galaxy. \citet*{HSD:1998} discussed the possibility 
that these were not the `usual' RSG variables, but possibly the evolved 
products of Hubble-Sandage variables, with pulsations driven by
Kelvin-Helmholtz instabilities as predicted by \citet{Heger:1997}.
In the case of the four objects in question here, V3, V15, V33 and V37,
the amplitudes, while larger 
than expected for RSG LPVs, are much smaller than the three stars mentioned 
in DDO\,187.  \citet{Heger:1997} expect periods near 900 days for the 
red descendants of Hubble-Sandage variables, if they are indeed driven by 
Kelvin-Helmholtz instabilities, as they predict, but this is not the case 
for the four stars at hand, which have periods from 210 to 820 days.
We must stress that the {\it empirical} behavior of RSG LPVs is not known 
very well, and the moderately larger amplitudes for the four 
objects in discussion may not be so unusual after all. Only a systematic
study of LPVs among young stellar populations with different metallicities 
can give us the answer.
 
% ******************************************************************
% 8. Summary
% ******************************************************************
\section{Summary}

We have presented the results of a search for variable stars in NGC~4395
using the WIYN 3.5 m and the KPNO 2.1 m telescopes. 
Since the observations have been accumulated over a time span of 8 years,
we have been able to not only discover 11 Cepheids, but also
37 long period variables 
(LPVs). A true distance modulus of $28.02 \pm 0.18$ and a 
mean reddening $E(g-r)$ of 0.06 and of $E(r-i)$ of 0.10 
have been derived from the 
apparent distance moduli in $g$, $r$ and $i$ based on the
LMC P-L relation by \citet*{Sandage:etal:2003}.
The de-reddened distance
modulus corresponds to a distance of $4.0 \pm 0.3$ Mpc. 
This adds to the growing list of galaxies well outside
the Local Group for which Cepheid distances are now
obtained from ground based observations. 
NGC~4395 is of particular interest, since
it is the nearest known Seyfert 1 galaxy.

Implying an upper limit on the luminosity of an AGB star of 
$M_{r} \approx -5.2$
\citep*{Kinman:Mould:Wood:1987} and a distance modulus of 28.02, we
concluded that 4 out of the 37 red long period variables 
are AGB stars, the rest probably beeing RSG LPVs. 

\smallskip\noindent
{\bf Acknowledgments}

The observations with the WIYN telescope were made in part
through the queue-scheduled service observing program that 
was beeing run by NOAO. We thank Dianne Harmer, Paul
Smith, and Daryl Willmarth for their participation
in the observations.

% ******************************************************************
%  End of ms.tex
% ******************************************************************

% ******************************************************************
% Bibliography
% ******************************************************************

% ******************************************************************
% Figure 
% ******************************************************************
% ******************************************************************

% ******************************************************************
% Figure 
% ******************************************************************
% ******************************************************************

% ******************************************************************
% Figure 
% ******************************************************************
\begin{figure}[!ht]
\epsscale{0.75}
\plotone{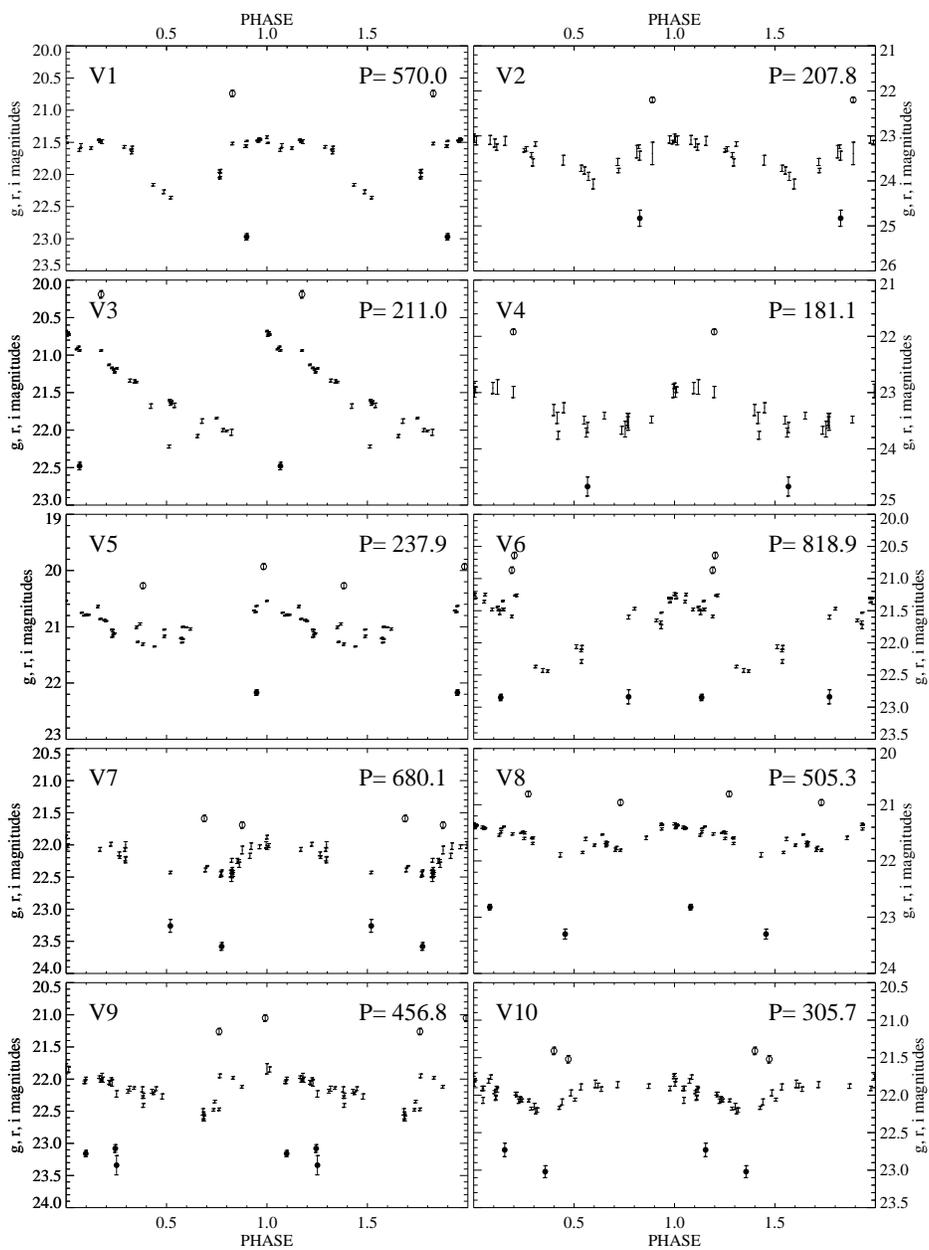}
\vspace{0cm}
\caption{The light curves of the Long Period Variables in $r$ (small 
dots with error bars), 
in $g$ (large filled circles) and in $i$ (large open circles). 
The identification numbers (V1-V37) and its periods
are displayed in each graph.
The periods
are given in days.}
\label{Thim.fig3a.eps}
\end{figure}

% ******************************************************************
% Figure 
% ******************************************************************
\begin{figure}[!ht]
\epsscale{0.75}
\plotone{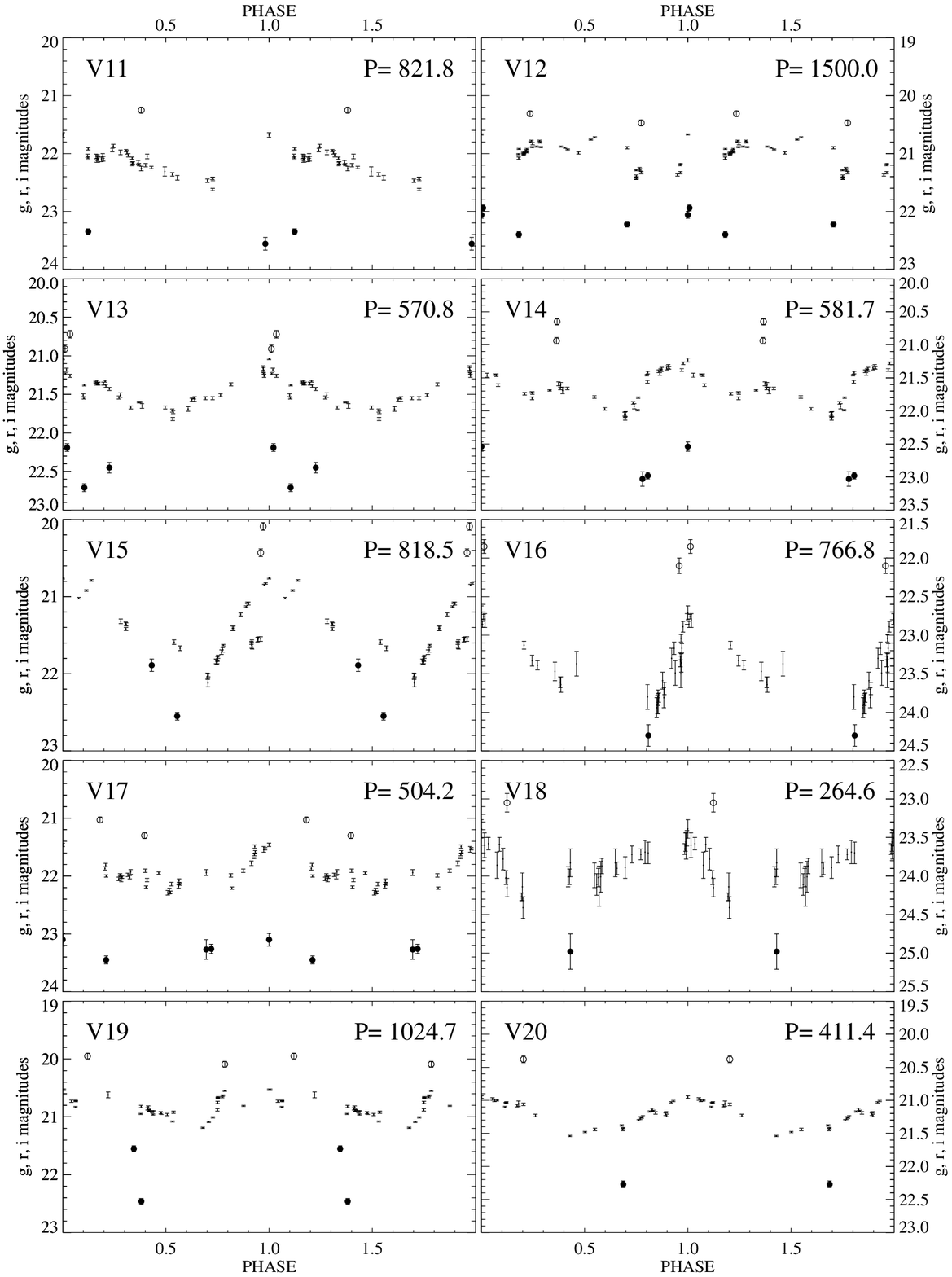}
\vspace{0cm}
\addtocounter{figure}{-1}
\caption{(contd.)}
\label{Thim.fig3b.eps}
\end{figure}

% ******************************************************************
% Figure 
% ******************************************************************
\begin{figure}[!ht]
\epsscale{0.75}
\plotone{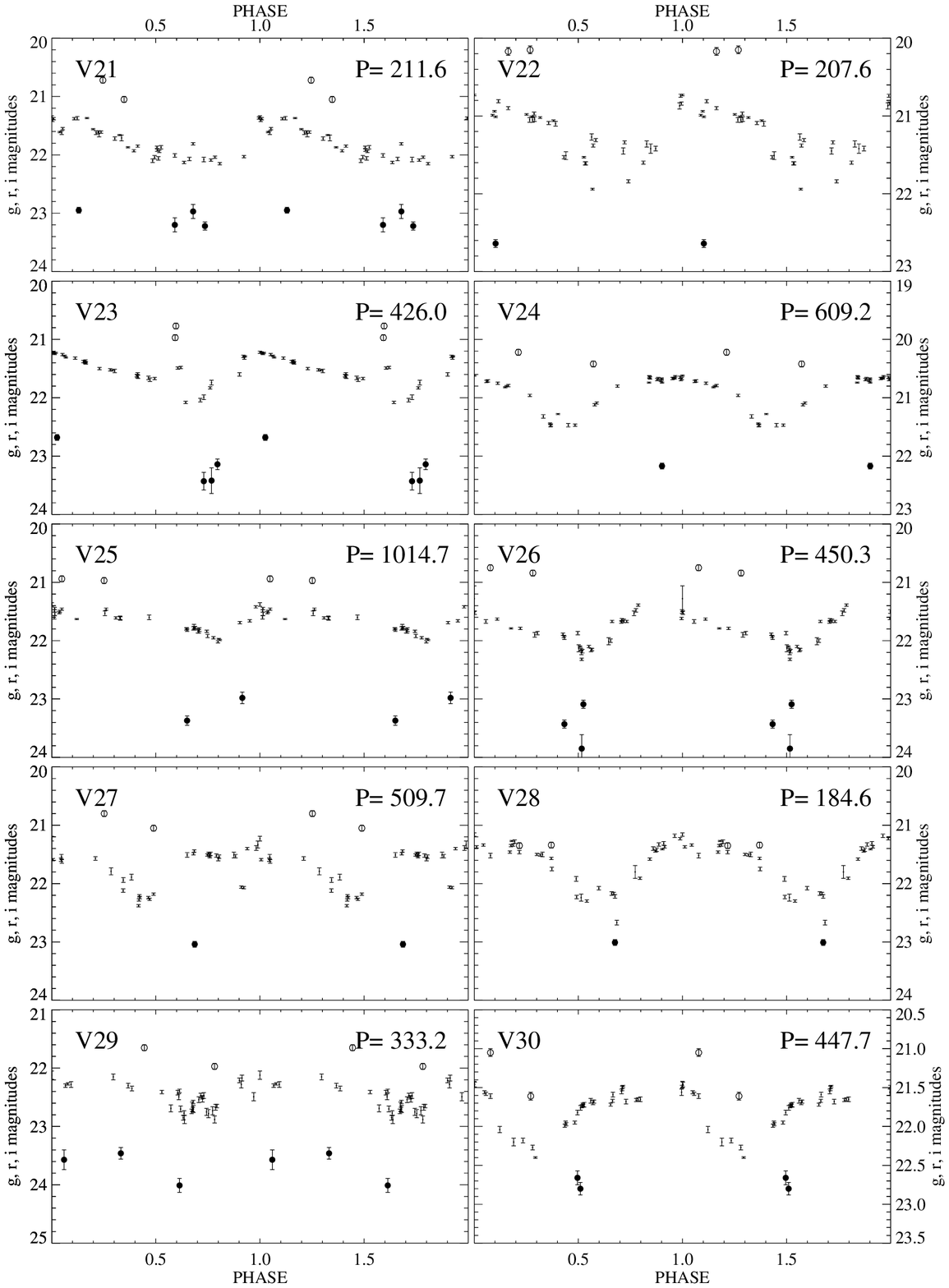}
\vspace{0cm}
\addtocounter{figure}{-1}
\caption{(contd.)}
\label{Thim.fig3c.eps}
\end{figure}

% ******************************************************************
% Figure 
% ******************************************************************
\begin{figure}[!ht]
\epsscale{0.75}
\plotone{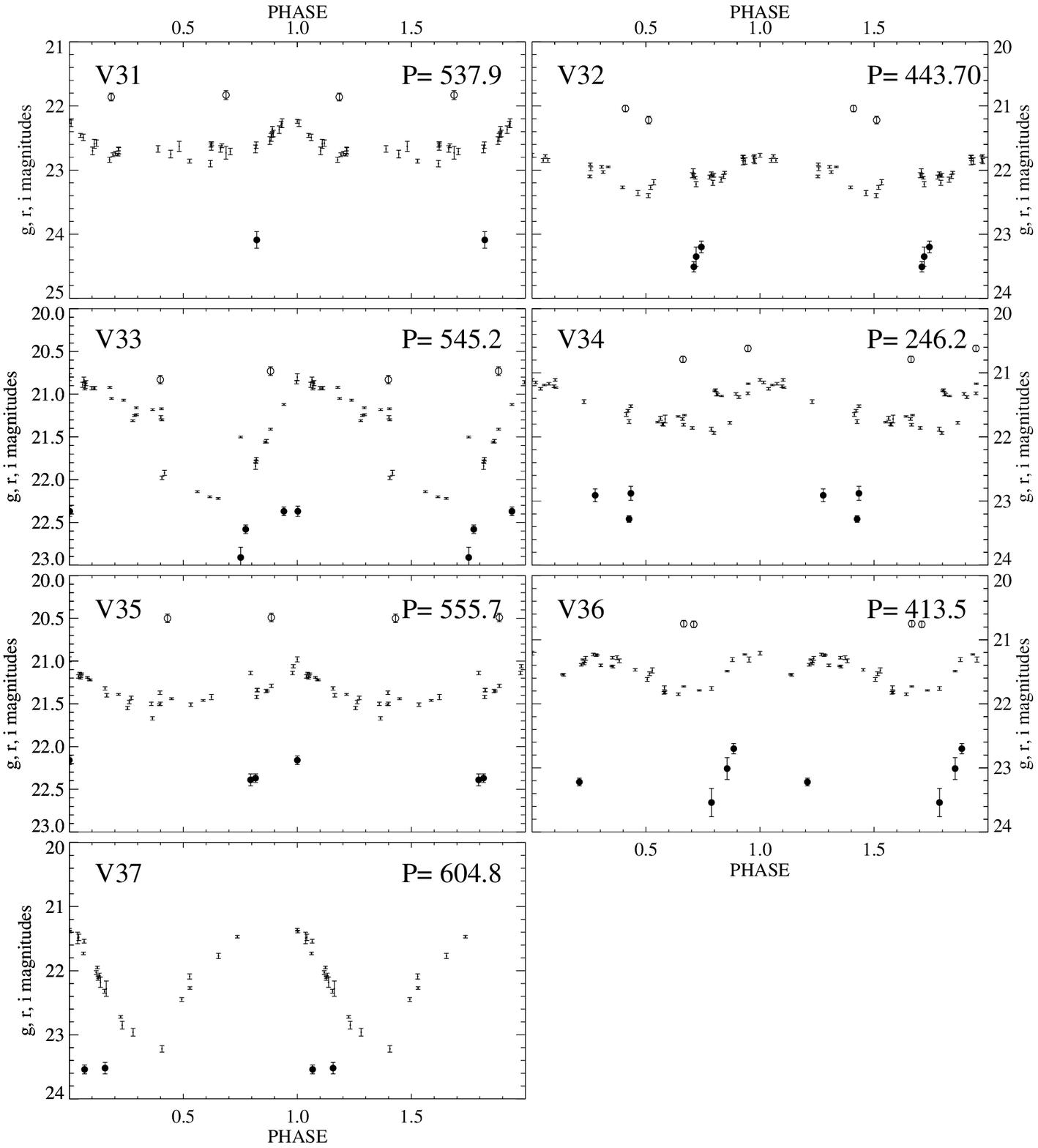}
\vspace{0cm}
\addtocounter{figure}{-1}
\caption{(contd.)}
\label{Thim.fig3d.eps}
\end{figure}
% *********

% ******************************************************************
% Figure 
% ******************************************************************
\begin{figure}[!ht]
\epsscale{0.7}
\plotone{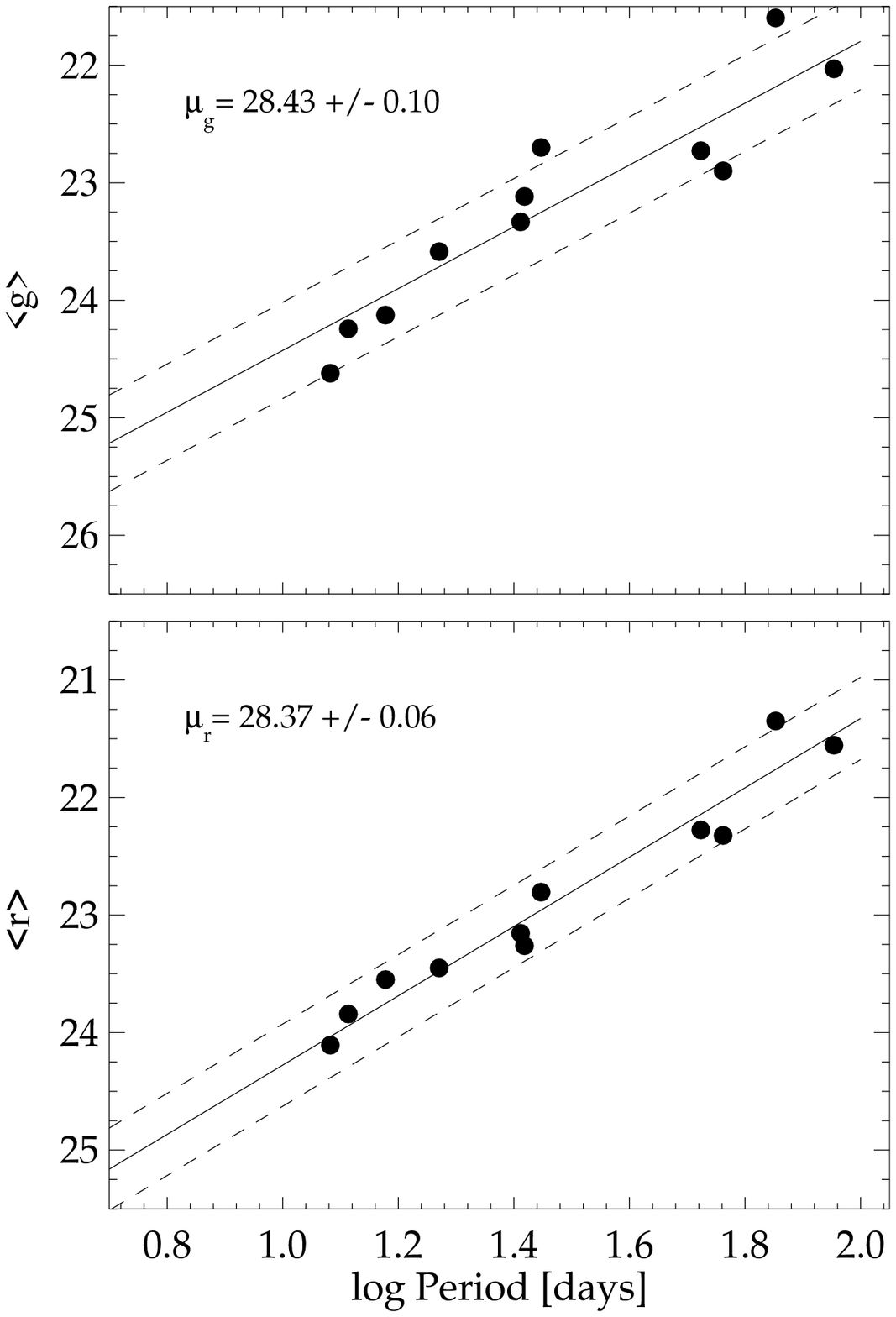}
\vspace{0cm}
\caption{Period-luminosity relation of NGC\,4395
in $g$ (top) and $r$ (bottom)
for all 11 Cepheids. The solid lines represent the best fit
with the slope of $-2.63$ in $g$ and $-2.95$ in $r$
from eq.~\ref{andy4} and~\ref{andy5}.
The dashed lines account for an adopted intrinsic width 
of the instability strip of $\pm$ 0.41 mag in $g$ and
$\pm$ 0.35 mag in $r$.}
\label{Thim.fig4.eps}
\end{figure}
% ******************************************************************

% ******************************************************************
% Figure 
% ******************************************************************
\begin{figure}[!ht]
\epsscale{0.55}
\plotone{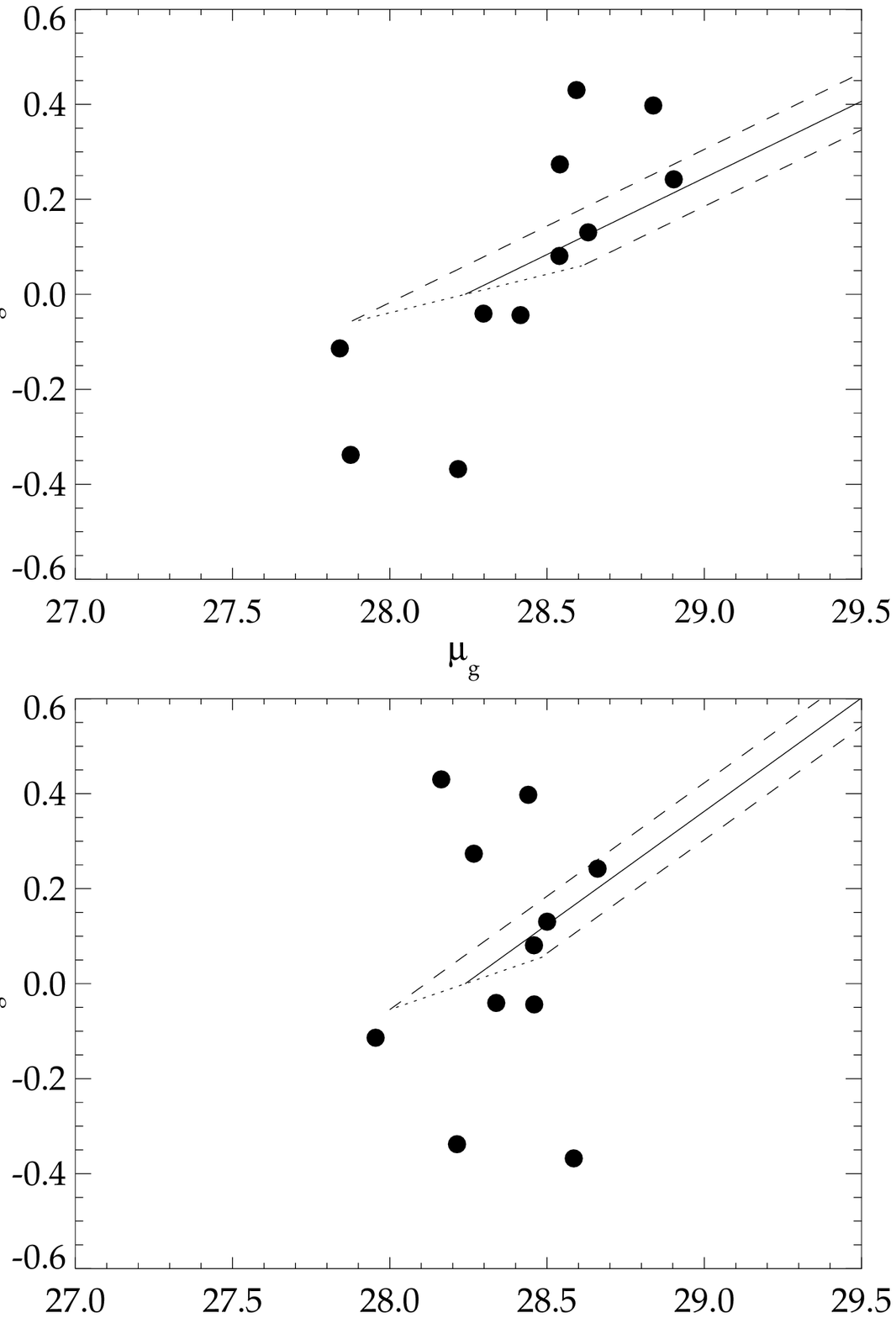}
\vspace{0cm}
\caption{Individual apparent distance moduli $\mu_g$ (upper panel)
and $\mu_r$ (lower panel) are plotted
against the difference in the apparent distance moduli in $g$
and $r$ for each Cepheid, which allows to detect the presence
of differential absorption. The bold lines indicates the reddening vectors,
the dashed lines the width of the instability strip. With
zero $E(g-r)$ reddening, the PL-relation is represented by a single point
at $\mu_g$ = $\mu_r$ = 28.22. Because of the width of the instability strip
this point is elongated and is shown as a dotted line in 
both figures. It is sloped due to the change of color
across the instability strip, for details
see \citet{Saha:etal:1996}. The observed points in the upper
panel are spread along a slope of 1 rather than on the reddening
line, in the lower panel the points seems to spread in 
vertical direction. The distribution of the points illustrates  
that they do not lie on the
reddening lines which indicates that there is no measurable
differential absorption at the level of the rms scatter in $\mu_r$.}
\label{Thim.fig5.eps}
\end{figure}
% ******************************************************************

% ******************************************************************
% Figure 
% ******************************************************************
\begin{figure}[!ht]
\epsscale{0.8}
\plotone{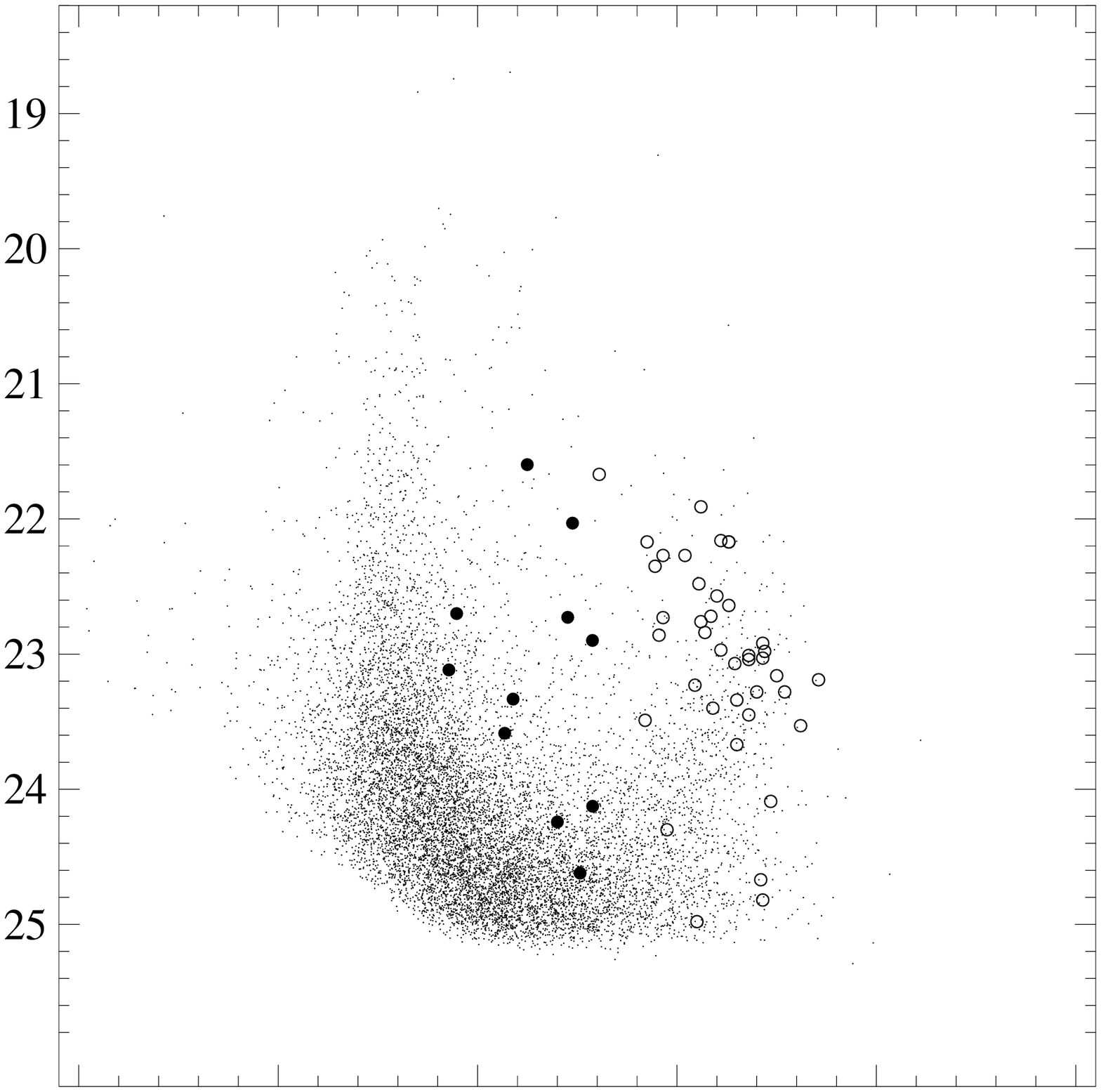}
\vspace{1cm}
\caption{$g-r$ versus $g$ color-magnitude diagram. Cepheids 
are plotted as filled circles, other variable candidates
are plotted as open circles.}
\label{Thim.fig6.eps}
\end{figure}
% ******************************************************************

% ******************************************************************
% Figure 
% ******************************************************************
\begin{figure}[!ht]
\epsscale{0.8}
\plotone{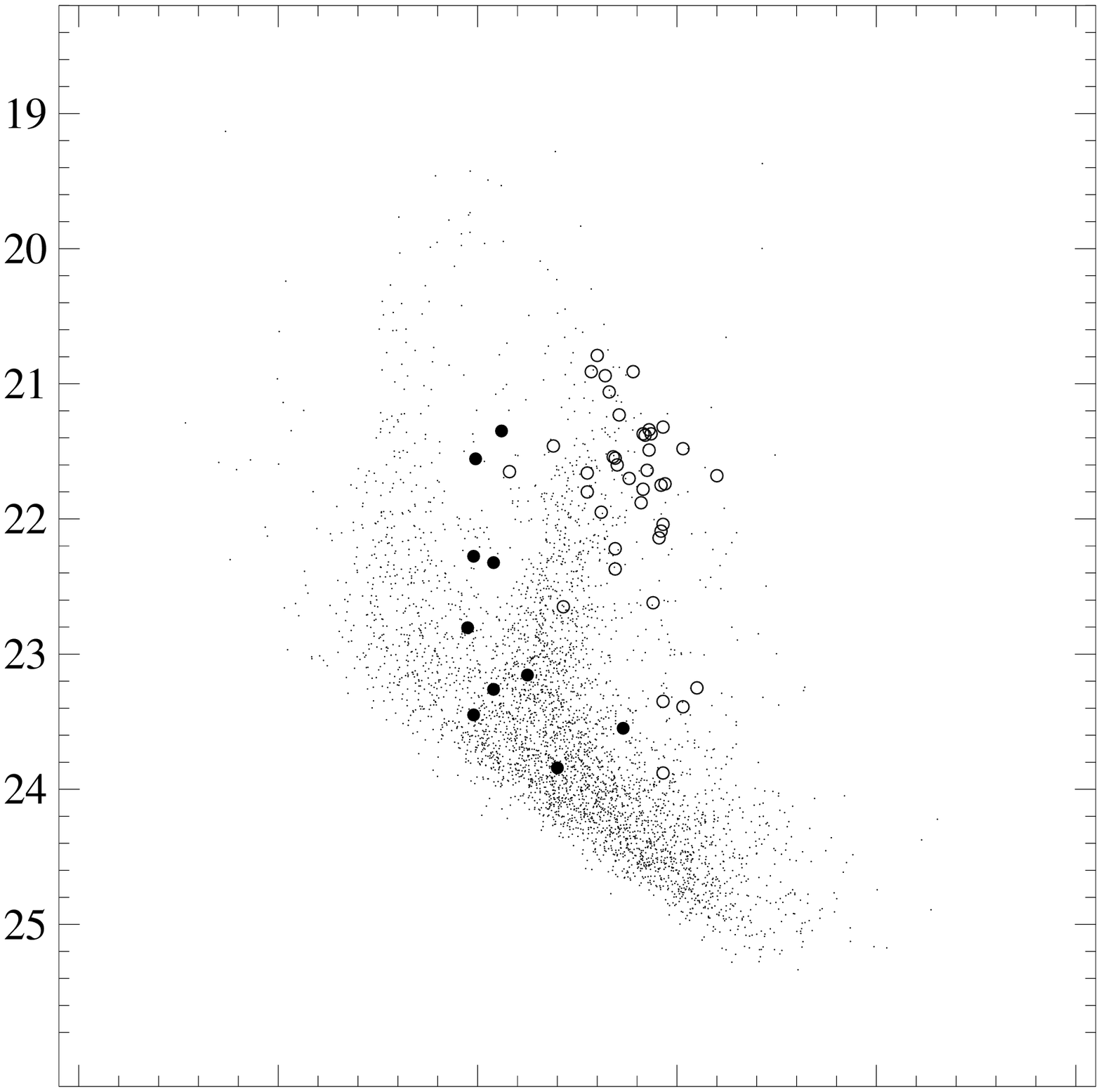}
\vspace{1cm}
\caption{$r-i$ versus $r$ color-magnitude diagram. Cepheids 
are plotted as filled circles, other variable candidates
are plotted as open circles.}
\label{Thim.fig7.eps}
\end{figure}
% ******************************************************************

% ******************************************************************
% Figure 
% ******************************************************************
\begin{figure}[!ht]
\epsscale{0.75}
\plotone{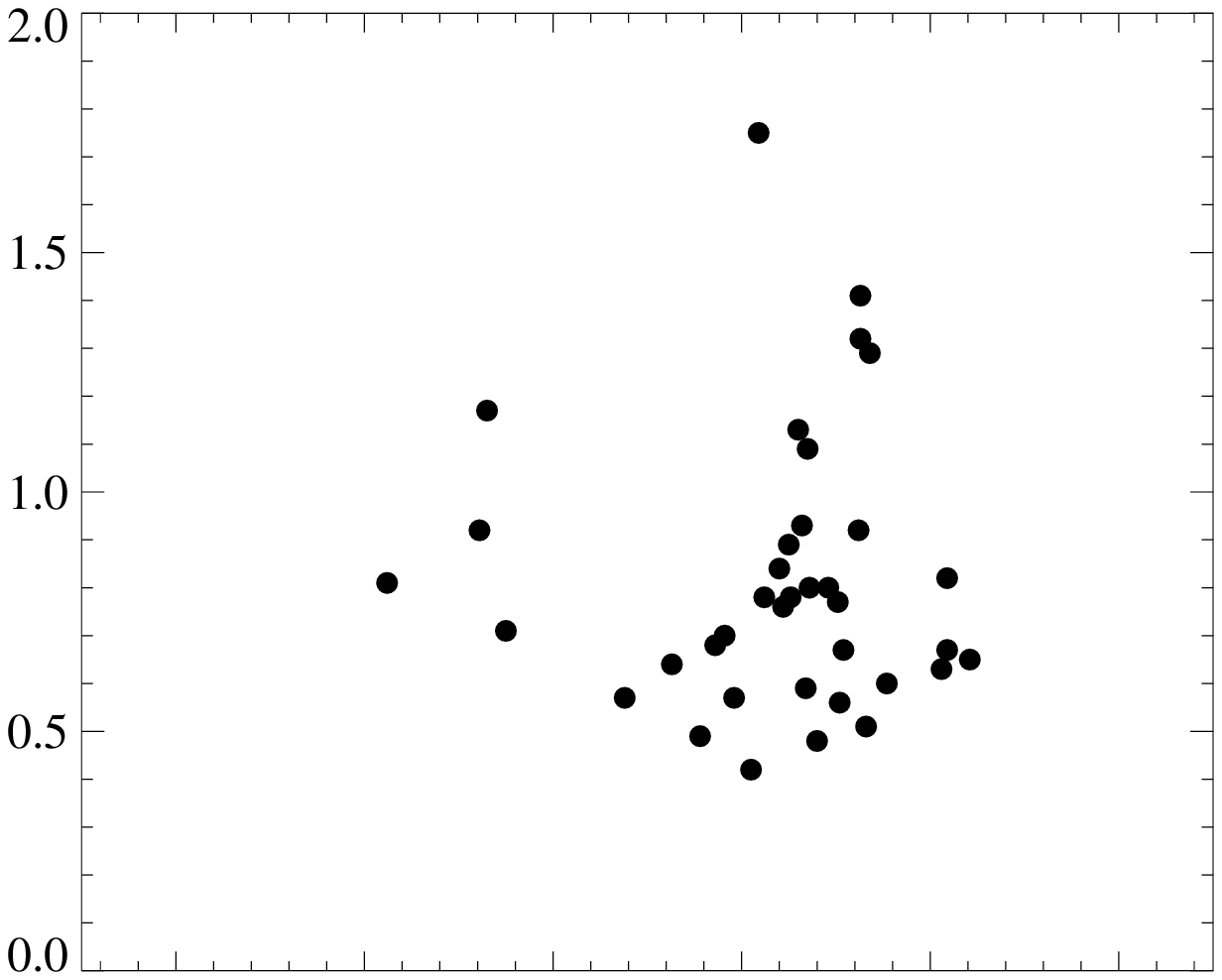}
\vspace{1cm}
\caption{Mean $r$ magnitudes for the long period variables
are plotted against their $r$ amplitude.}
\label{Thim.fig8.eps}
\end{figure}
% ******************************************************************

\newpage
\clearpage
\pagebreak

%%%%%%%%
%%%%%%%%%%%%    TABLE 1     %%%%%%%%%%%%%%%%%%%%%
%%%%%%%
\begin{deluxetable}{crccccc}
\tablecaption{Journal of Observations
\label{tab1}}
\tablewidth{0pt}
\tablehead{
\colhead{Epoch ($HJD$)\tablenotemark{a}} & \colhead{Exposure} & \colhead{Filter} & \colhead{Seeing} & \colhead{Telescope} & \colhead{Instrument} \\
\colhead{} & \colhead{time [s]} & \colhead{} & \colhead{[arc-sec]} & \colhead{} & \colhead{}}
\startdata
 48269.94 & 4 x 900  & r & 1.0   & KP 2.1m & T1KA \\
 48359.75 & 3 x 900  & r & 1.0   & KP 2.1m & TI-2 \\
 48388.77 & 3 x 1200 & r & 1.0   & KP 2.1m & TI-2 \\
 48622.95 & 3 x 900  & r & 1.3   & KP 2.1m & T1KA \\
 48622.95 & 2 x 900  & g & 1.2   & KP 2.1m & T1KA \\
 49064.83 & 3 x 1200 & r & 1.0   & KP 2.1m & T1KA \\
 49064.83 & 1 x 1200 & g & 1.0   & KP 2.1m & T1KA \\
 49076.86 & 3 x 900  & g & 1.2   & KP 2.1m & T1KA \\
\enddata
\tablenotetext{a}{HJD-2400000}
\end{deluxetable}
%
%%%%%%%
{
\begin{deluxetable}{crrcccrr}
\tablecaption{Local Standards}
\tablewidth{0pt}
\tablehead{
\colhead{Identifier} &
\colhead{$X$} &
\colhead{$Y$} &
\colhead{$g$} &
\colhead{$r$} &
\colhead{$i$} &
\colhead{RA [J2000]} &
\colhead{DEC [J2000]} 
}
\startdata
A  & 1561 &  513 & 17.59 & 16.97 & 16.76 & 12:25:54.2 & +33:31:40 \\
B  & 1337 &  851 & 18.69 & 18.13 & 17.93 & 12:25:57.5 &  33:32:47 \\
C  & 1116 &  850 & 18.49 & 18.20 & 18.09 & 12:26:01.2 &  33:32:48 \\
D  &  694 & 1024 & 18.54 & 17.41 & 16.36 & 12:26:08.1 &  33:33:24 \\
Z  & 1746 &  444 & 16.03 & 15.81 & 15.73 & 12:25:51.2 &  33:31:26 \\
f1 &  472 & 1468 & 21.77 & 20.55 & 20.10 & 12:26:11.2 &  33:34:51 \\
f2 &  148 &  972 & 20.94 & 20.12 & 19.78 & 12:26:17.3 &  33:33:16 \\
f3 & 1210 & 1422 & 20.53 & 19.39 & 18.00 & 12:25:58.9 &  33:34:39 \\ 
\enddata
\label{tab_stds}
\end{deluxetable}
}

%%%%%
%%%%%%%%%%%%%%%%%%%%   Table    %%%%%%%%%%%%%%%%%%%
%%%%%%%
%%%%%
%
\begin{deluxetable}{ccccccccc}
\tabletypesize{\tiny}
\tablecaption{Photometry of the Cepheids: Magnitudes and Error Estimates
\label{tab_phot_cep}}
\tablewidth{0pt}
\tablehead{ }
\startdata
\tableline\tableline
 HJD & C1 & C2 & C3 & C4 & C5 & C6 & C7 & C8 \\
\tableline
\multicolumn{9}{c}{$r$} \\
\tableline
 2448269.9406 & \nodata \nodata & 23.81  0.16  & \nodata \nodata & 21.30  0.03 &  23.98  0.18 &  23.15  0.10 &  23.23  0.08  & \nodata \nodata   \\
 2448359.7523 & \nodata \nodata & \nodata \nodata & \nodata \nodata & \nodata \nodata & 23.27  0.10 & \nodata \nodata & \nodata \nodata & \nodata \nodata \\
 2448388.7788 & \nodata \nodata & \nodata \nodata & \nodata \nodata & \nodata \nodata & 23.18  0.12 & \nodata \nodata & \nodata \nodata  & \nodata \nodata \\
 2448622.9552 & \nodata \nodata & \nodata \nodata & 23.86  0.18 &  21.30  0.02 &  23.22  0.10 &  22.83  0.25 &  22.92  0.09  & \nodata \nodata   \\
 2449064.8339 & \nodata \nodata  & \nodata \nodata & \nodata \nodata & 21.39  0.02 &  22.86  0.07 &  22.93  0.08 &  22.62  0.08  & \nodata \nodata   \\
 2449421.8408 & \nodata \nodata & 24.12  0.14 &  24.30  0.17 &  21.39  0.02  & \nodata \nodata & 22.86  0.05 &  22.31  0.03 &  24.05  0.19   \\
\enddata
\end{deluxetable}
%

%%%%%
%%%%%%%%%%%%%%%%%%%%   Table    %%%%%%%%%%%%%%%%%%%
%%%%%%%
%%%%%
%
\begin{deluxetable}{ccccccccc}
\tabletypesize{\tiny}
\tablecaption{Photometry of the Variable Stars: Magnitudes and Error Estimates
\label{tab_phot_var}}
\tablewidth{0pt}
\tablehead{ }
\startdata
\tableline\tableline
 HJD & V1 & V2 & V3 & V4 & V5 & V6 & V7 & V8 \\
\tableline
\multicolumn{9}{c}{$r$} \\
\tableline
 2448269.9406 & \nodata \nodata & \nodata \nodata & \nodata \nodata & \nodata \nodata & 20.64  0.02 & 21.47  0.02 & 21.88  0.03 & 21.59  0.03 \\
 2448359.7523 & \nodata \nodata & \nodata \nodata & \nodata \nodata & \nodata \nodata & \nodata \nodata & 21.65  0.02 & \nodata \nodata & \nodata \nodata \\
 2448388.7788 & \nodata \nodata & \nodata \nodata & \nodata \nodata & \nodata \nodata & \nodata \nodata & \nodata \nodata & \nodata \nodata & \nodata \nodata \\
 2448622.9552 & \nodata \nodata & \nodata \nodata & \nodata \nodata & \nodata \nodata & \nodata \nodata & \nodata \nodata & 22.43  0.06 & 21.61  0.03 \\
 2449064.8339 & \nodata \nodata & \nodata \nodata & \nodata \nodata & \nodata \nodata & \nodata \nodata & 21.60  0.03 & 22.07  0.04 & 21.89  0.04 \\
 2449421.8408 & \nodata \nodata & \nodata \nodata & \nodata \nodata & \nodata \nodata & 20.55  0.01 & 21.26  0.01 & 22.39  0.03 & 21.42  0.01 \\
\enddata
\end{deluxetable}
%

%%%%%%%%
%%%%%%%%%%%%    TABLE      %%%%%%%%%%%%%%%%%%%%%
%%%%%%%
\vspace{-4cm}

{
\begin{deluxetable}{ccccccccrrcc}
\rotate
\tablecaption{Cepheid Variables}
\tablewidth{0pt}
\vspace{-4cm}
\tablehead{
\colhead{Object} &
\colhead{Period} &
\colhead{$<\!g\!>$\tablenotemark{a}} &
\colhead{$\sigma_{< g >}$} &
\colhead{$<\!r\!>$\tablenotemark{b}} &
\colhead{$\sigma_{< r >}$} &
\colhead{$<\!i\!>$\tablenotemark{c}} &
\colhead{$\sigma_{< i >}$} &
\colhead{$X$} & 
\colhead{$Y$} &
\colhead{RA [J2000] } &
\colhead{DEC[J2000] } \\
\colhead{} &
\colhead{[days]} &
\colhead{} &
\colhead{} &
\colhead{} &
\colhead{} &
\colhead{} & 
\colhead{} &
\colhead{} }
\startdata
C01 & 18.65 & 23.59 & 0.21 & 23.45  & 0.09 & 23.47 & 0.24 &  177.17 & 931.06    &  12:26:16.8 & 33:33:07 \\
C02 & 12.99 & 24.24 & 0.26 & 23.84  & 0.11 & 23.44 & 0.23 &  477.19 & 143.28    &  12:26:12.9 & 33:30:33 \\
C03 & 15.06 & 24.13 & 0.25 & 23.55  & 0.10 & 22.82 & 0.19 &  863.10 & 198.74    &  12:26:06.3 & 33:30:42 \\
C04 & 71.26 & 21.60 & 0.08 & 21.35  & 0.02 & 21.23 & 0.16 &  870.34 & 152.16    &  12:26:06.3 & 33:30:33 \\
C05 & 26.20 & 23.12 & 0.19 & 23.26  & 0.08 & 23.18 & 0.21 &  962.14 & 780.46    &  12:26:03.9 & 33:32:35 \\
C06 & 25.80 & 23.33 & 0.19 & 23.16  & 0.08 & 22.91 & 0.18 & 1493.71 & 878.36   &  12:25:54.9 & 33:32:51 \\
C07 & 27.99 & 22.70 & 0.30 & 22.81  & 0.06 & 22.86 & 0.22 & 1551.86 & 69.44    &  12:25:55.0 & 33:30:14 \\
C08 & 12.09 & 24.62 & 0.29 & 24.11  & 0.17 & \nodata &\nodata & 1567.74 & 346.60 &  12:25:54.3 & 33:31:07 \\
C09 & 57.80 & 22.90 & 0.05 & 22.32  & 0.04 & 22.24 & 0.16 & 1612.33 & 419.98   &  12:25:53.5 & 33:31:22 \\
C10 & 52.90 & 22.73 & 0.13 & 22.28  & 0.04 & 22.30 & 0.17 & 1647.97 & 1365.20  &  12:25:51.6 & 33:34:25 \\
C11 & 89.89 & 22.03 & 0.03 & 21.56  & 0.02 & 21.57 & 0.16 & 1810.24 & 1237.91  &  12:25:49.1 & 33:34:00 \\
\enddata
\tablenotetext{a}{Mean $g$ as computed in equation \ref{eq_labhardt_mean}}
\tablenotetext{b}{Phase-weighted intensity average}
\tablenotetext{c}{Mean $i$ from $r-i$ at $i$ phase}
\label{tab_cepheids}
\end{deluxetable}
}

%%%%%%%%
%%%%%%%%%%%%    TABLE      %%%%%%%%%%%%%%%%%%%%%
%%%%%%%
{
\begin{deluxetable}{ccccccrrcc}
\tablecaption{Other Variable Stars}
\tablewidth{0pt}
\tablehead{
\colhead{Object} &
\colhead{Period} &
\colhead{$<\!g\!>$\tablenotemark{a}} &
\colhead{$<\!r\!>$\tablenotemark{a}} &
\colhead{$<\!i\!>$\tablenotemark{b}} &
\colhead{$\Delta r$\tablenotemark{c}} &
\colhead{$X$} & 
\colhead{$Y$} &
\colhead{RA [J2000] } &
\colhead{DEC[J2000] } \\
\colhead{} &
\colhead{[days]} &
\colhead{} &
\colhead{} &
\colhead{} &
\colhead{} &
\colhead{} & 
\colhead{} &
\colhead{} &
\colhead{}}
\startdata
V01 & 570.0\tablenotemark{e} & 22.97 &  21.75  & 20.83  & 0.89    & 144.09 & 1154.40 &    12:26:17.1 & 33:33:51 \\
V02 &	207.8 & 24.82 &  23.39  & 22.36                   & 0.92  & 208.96 & 360.19 & 	  12:26:17.1 & 33:31:16 \\
V03 &	211.0\tablenotemark{e} & 22.48 &  21.37  & 20.54  & 1.32  & 216.63 & 202.95 & 	  12:26:17.1 & 33:30:46 \\
V04 &	181.1 & 24.67 &  23.25  & 22.15                   & 0.71  & 336.22 & 1638.35 & 	  12:26:13.2 & 33:35:24 \\
V05 &	237.9\tablenotemark{e} & 22.17 &  20.91  & 20.13  & 0.67  & 515.87 & 322.67 & 	  12:26:12.0 & 33:31:07 \\
V06 &	818.9\tablenotemark{e} & 22.84 &  21.70  & 20.94  & 1.13  & 651.49 & 536.32 & 	  12:26:09.4 & 33:31:48 \\
V07 &	680.1 & 23.40 &  22.22  & 21.53                   & 0.49  & 731.36 & 339.72 & 	  12:26:08.3 & 33:31:10 \\
V08 &	505.3 & 23.03 &  21.60  & 20.90                   & 0.48  & 953.77 & 282.99 & 	  12:26:04.7 & 33:30:58 \\
\enddata
\tablenotetext{a}{Phase-weighted intensity average}
\tablenotetext{b}{Mean $i$ from $r-i$ at $i$ phase}
\tablenotetext{c}{$\Delta r$ = $r$ peak to peak amplitude}
\tablenotetext{e}{Period uncertain}
\label{tab_variables}
\end{deluxetable}
}

\begin{deluxetable}{ccccc|cccc}
\tablecaption{Individual distance moduli and reddening values %$E(V$$-$$I)$
of Cepheids in NGC\,4395 using $g$ and $r$ magnitudes. }
\tablewidth{0pt}
\tablehead{
\colhead{} &
\multicolumn{4}{c}{PL-relation with trans. by Kent$^{\rm \, a}$} &
\multicolumn{4}{c}{PL-relation with synthetic trans.$^{\rm \, b}$} \\
\colhead{ID}  &  
\colhead{$\mu_g$} &
\colhead{$\mu_r$} &
\colhead{$E (g$$-$$r$)} &
\colhead{$\mu_0$} &
\colhead{$\mu_g$} &
\colhead{$\mu_r$} &
\colhead{$E (g$$-$$r$)} &
\colhead{$\mu_0$}}
\startdata
C01 & 28.375 & 28.278 & 0.097 & 28.040 & 28.298 & 28.339 &-0.041 & 28.438 \\  
C02 & 28.620 & 28.213 & 0.407 & 27.219 & 28.541 & 28.267 & 0.273 & 27.600 \\  
C03 & 28.672 & 28.107 & 0.565 & 26.727 & 28.594 & 28.164 & 0.430 & 27.114 \\   
C04 & 27.913 & 27.871 & 0.042 & 27.769 & 27.841 & 27.955 &-0.114 & 28.233 \\   
C05 & 28.293 & 28.518 &-0.225 & 29.068 & 28.217 & 28.585 &-0.368 & 29.483 \\   
C06 & 28.491 & 28.393 & 0.099 & 28.152 & 28.416 & 28.459 &-0.044 & 28.566 \\   
C07 & 27.951 & 28.146 &-0.195 & 28.621 & 27.876 & 28.214 &-0.338 & 29.039 \\   
C08 & 28.917 & 28.387 & 0.530 & 27.093 & 28.838 & 28.440 & 0.398 & 27.470 \\   
C09 & 28.975 & 28.580 & 0.395 & 27.616 & 28.903 & 28.661 & 0.242 & 28.070 \\   
C10 & 28.703 & 28.421 & 0.282 & 27.733 & 28.631 & 28.500 & 0.130 & 28.182 \\   
C11 & 28.609 & 28.370 & 0.239 & 27.786 & 28.539 & 28.459 & 0.081 & 28.261 \\ \hline    
Mean: & 28.502 & 28.299 & 0.203 & 27.802 & 28.427 & 28.367 & 0.059 & 28.223  \\ 
      & $\pm$0.104&$\pm$0.061&$\pm$0.081&$\pm$0.202&$\pm$0.104&$\pm$0.062&$\pm$0.081 &$\pm$0.205\\
\enddata
\tablenotetext{a}{PL-relation of \cite{Madore:Freedman:1991} with 
transformations by \cite{Kent:1985}.}
\tablenotetext{b}{PL-relation of \cite{Madore:Freedman:1991} with
synthetic transformations.}
\label{tab_individual}
\end{deluxetable}

\begin{deluxetable}{ccccc|cccc}
\tablecaption{Individual distance moduli and reddening values %$E(V$$-$$I)$
of Cepheids in NGC\,4395 using $r$ and $i$ magnitudes. }
\tablewidth{0pt}
\tablehead{
\colhead{} &

\multicolumn{4}{c}{PL-relation with trans. by Kent$^{\rm \, a}$} &
\multicolumn{4}{c}{PL-relation with synthetic trans.$^{\rm \, b}$} \\
\colhead{ID}  &  
\colhead{$\mu_r$} &
\colhead{$\mu_i$} &
\colhead{$E (r$$-$$i$)} &
\colhead{$\mu_0$} &
\colhead{$\mu_r$} &
\colhead{$\mu_i$} &
\colhead{$E (r$$-$$i$)} &
\colhead{$\mu_0$}}
\startdata
C01 & 28.278 & 28.342 &  -0.064 & 28.517 & 28.339 & 28.381 &  -0.042 & 28.497 \\  
C02 & 28.213 & 27.841 &   0.372 & 26.830 & 28.267 & 27.869 &   0.398 & 26.785  \\ 
C03 & 28.107 & 27.414 &   0.693 & 25.529 & 28.164 & 27.446 &   0.718 & 25.494 \\  
C04 & 27.871 & 27.849 &   0.022 & 27.788 & 27.955 & 27.928 &   0.027 & 27.856  \\ 
C05 & 28.518 & 28.495 &   0.023 & 28.431 & 28.585 & 28.544 &   0.041 & 28.433  \\ 
C06 & 28.393 & 28.205 &   0.188 & 27.694 & 28.459 & 28.254 &   0.206 & 27.694 \\  
C07 & 28.146 & 28.261 &  -0.115 & 28.574 & 28.214 & 28.312 &  -0.099 & 28.581  \\ 
C08 & 28.387 &\nodata & \nodata &\nodata & 28.440 & \nodata & \nodata & \nodata \\ 
C09 & 28.580 & 28.586 &  -0.006 & 28.601 & 28.661 & 28.659 &   0.002 & 28.655  \\ 
C10 & 28.421 & 28.530 &  -0.109 & 28.827 & 28.500 & 28.601 &  -0.101 & 28.875  \\ 
C11 & 28.370 & 28.491 &  -0.121 & 28.820 & 28.459 & 28.578 &  -0.119 & 28.903  \\ 
Mean: & 28.299 & 28.201 & 0.088 & 27.961 & 28.367 & 28.257 & 0.103 & 27.977  \\ 
      & $\pm$0.061&$\pm$0.121&$\pm$0.083&$\pm$0.335&$\pm$0.062&$\pm$0.125&$\pm$0.085 &$\pm$0.344\\
\enddata
\tablenotetext{a}{PL-relation of \cite{Madore:Freedman:1991} with 
transformations by \cite{Kent:1985}.}
\tablenotetext{b}{PL-relation of \cite{Madore:Freedman:1991} with
synthetic transformations.}
\label{tab_individual_r_i}
\end{deluxetable}


\begin{thebibliography}{}
%
\bibitem[Burstein \& Heiles(1984)]{Burstein:Heiles:1984} 
Burstein, D.~\& 
Heiles, C.\ 1984, \apjs, 54, 33 
%
\bibitem[Cardelli, Clayton, \& Mathis(1989)]{Cardelli:Clayton:Mathis:1989}
Cardelli, J.A., Clayton, G.C., \& Mathis, J.S. 1989, 
ApJ, 345, 245
%
\bibitem[de Vaucouleurs(1975)]{de Vaucouleurs:1975}
de Vaucouleurs, G.\ 1975, Social Studies of Science, 9, 557 
%
\bibitem[Filippenko \& Ho(2003)]{Filippenko:Ho:2003}
Filippenko, A.~V., \& Ho, L.~C. 2003, \apjl, 588, 13
%
\bibitem[Freedman(1988)]{Freedman:1988}
Freedman, W.~L.\ 1988, \apj, 326, 691 
%
\bibitem[Heger et~al.(1997)]{Heger:1997} Heger, A., Jeannin, L., Langer, N., 
\& Baraffe, I. \ 1997, A\&A 327, 224
%
\bibitem[Hoessel et~al.(1994)]{Hoessel:etal:1994}
Hoessel, J.~G., Saha, A., Krist, J., \& Danielson, G.~E.\ 1994, 
\aj, 108, 645 
%
\bibitem[Hoessel, Saha, \& Danielson(1998)Hoessel et~al.]{HSD:1998} Hoessel, J.~G., 
Saha, A., \& Danielson, G.~E.\ 1998, \aj, 116, 1687
%
\bibitem[Hubble \& Sandage(1953)]{Hubble:Sandage:1953}
Hubble, E.~\& Sandage, A.\ 1953, \apj, 118, 353 
%
\bibitem[Karachentsev \& Drozdovsky(1998)]{Karachentsev:Drozdovsky:1998} 
Karachentsev, I.~D.~\& Drozdovsky, I.~O.\ 1998, \aaps, 131, 1 
%
\bibitem[Karachentsev et~al.(2003)]{Karachentsev:etal:2003} 
Karachentsev, I.~D.~et al.\ 2003, \aap, 398, 467 
%
\bibitem[Kent(1985)]{Kent:1985}
   Kent, S. M. 1985, \pasp, 97, 165
%
\bibitem[Kholopov et~al.(1985)]{Kholopov:etal:1985}
Kholopov, P. N., Samus, N. N., Frolov, M. S., et al. 1985-88, 
General Catalouge of Variable Stars. 4th edition, 
Nauka Publishing House, Moscow (GCVS) 
%
\bibitem[Kinman, Mould, \& Wood(1987)Kinman et~al.]{Kinman:Mould:Wood:1987}
Kinman, T.~D., 
Mould, J.~R., \& Wood, P.~R.\ 1987, \aj, 93, 833 
%
\bibitem[Kraan-Korteweg \& Tammann(1979)]{Kraan-Korteweg:Tammann:1979}
   Kraan-Korteweg, R.C., \& Tammann, G.A. 1979,
   AN, 300, 181
%
\bibitem[Kurucz()]{Kurucz}
Kurucz, http://kurucz.harvard.edu/
%
\bibitem[Labhardt et~al.(1997)Labhardt, Sandage, \& Tammann]{Labhardt:etal:1997}
   Labhardt, L., Sandage, A., \& Tammann, G.A. 1997,
   A\&A, 322, 751
%
\bibitem[Lafler \& Kinman(1965)]{Lafler:Kinman:1965}
   Lafler, J., \& Kinman, T.D. 1965,
   \apjs, 11, 216
%
\bibitem[Madore \& Freedman(1991)]{Madore:Freedman:1991}
   Madore, B.F., \& Freedman, W.L. 1991,
   \pasp, 103, 933
%
\bibitem[Minitti et~al.(2003), in preparation]{Minitti:etal:2003}
Minitti et~al., 2003, in preparation
%
\bibitem[Rowan-Robinson(1985)]{Rowan:Robinson:1985}
Rowan-Robinson, M.\ 1985, New York, W.~H.~Freeman and Co, 1985, 364 
%
\bibitem[Roy et al.(1996)]{Roy:etal:1996}
Roy, J., Belley, J., Dutil, Y., \& Martin, P.\ 1996, \apj, 460, 284 
%
\bibitem[Saha \& Hoessel(1990)]{Saha:Hoessel:1990}
   Saha, A., \& Hoessel, J.G. 1990,
   \aj, 99, 97
%
\bibitem[Saha et al.(1996)]{Saha:etal:1996}
Saha, A., Sandage, A., 
Labhardt, L., Tammann, G.~A., Macchetto, F.~D., \& Panagia, N.\ 1996, \apj, 
466, 55 
%
\bibitem[Saha, Claver, \& Hoessel(2002)Saha et~al.]{Saha:Claver:Hoessel:2002}
Saha, A., Claver, J., \& Hoessel, J.~G.\ 2002, \aj, 124, 839 
%
\bibitem[Sandage \& Bedke(1994)]{Sandage:Bedke:1994}
Sandage, A.~\& Bedke, J.\ 1994, Washington, 
DC: Carnegie Institution of Washington with The 
Flintridge Foundation, |c1994
%
\bibitem[Sandage, Reindl \& Tammann(2003)Sandage et~al.]{Sandage:etal:2003}
   Sandage, A., \& Reindl, B., \& Tammann, G.A. 2003,
   in preparation
%
\bibitem[Schechter, Mateo, \& Saha(1993)]{Schechter:Mateo:Saha:1993}
Schechter, P.L., Mateo, M.L., \& Saha, A. 1993,
\pasp, 105, 1342
%
\bibitem[Schlegel, Finkbeiner, \& Davis(1998)]{Schlegel:Finkbeiner:Davis:1998} 
Schlegel, D.~J., Finkbeiner, D.~P., \& Davis, M.\ 1998, \apj, 500, 525 
%
\bibitem[Tammann et al.(2002)]{Tammann:etal:2002} 
Tammann, G.~A., Reindl, B., Thim, F., Saha, A., \& Sandage, A.\ 
2002, ASP Conf.~Ser.~283: A New Era in Cosmology, 258 
%
\bibitem[Tammann, Sandage, \& Reindl(2003)Tammann et~al.]{Tammann:Sandage:Reindl:2003}
Tammann, G.~A., Sandage, A., \& Reindl, B.\ 2003, \aap, 404, 423 
%
\bibitem[Thuan \& Gunn(1976)]{Thuan:Gunn:1976}
Thuan, T.~X.~\& Gunn, 
J.~E.\ 1976, \pasp, 88, 543 
%
\bibitem[Udalski et al.(1999)]{Udalski:etal:1999} 
Udalski, A., Szymanski, M., Kubiak, M., Pietrzynski, G., Soszynski, 
I., Wozniak, P., \& Zebrun, K.\ 1999, Acta Astronomica, 49, 201 
%
\bibitem[Wade et al.(1979)]{Wade:etal:1979} 
Wade, R.~A., Hoessel, J.~G., Elias, J.~H., \& Huchra, J.~P.\ 1979, \pasp, 
91, 35 
%
\bibitem[Wood \& Bessell(1985)]{Wood:Bessel:1985}
Wood, P.~R.~\& Bessell, M.~S.\ 1985, \pasp, 97, 681 
%
\bibitem[Wood, Bessell, \& Fox(1983)]{Wood:Bessell:Fox:1983}
Wood, P.~R., Bessell, M.~S., \& Fox, M.~W.\ 1983, \apj, 272, 99 
%
\end{thebibliography}
\end{document}